\documentclass[12pt,preprint]{aastex}

\slugcomment{submitted to AJ}

\shorttitle{VLA-COSMOS Survey}
\shortauthors{Schinnerer, Carilli et al.}

\begin{document}

\title{The VLA-COSMOS Survey: \\
    I. Radio Identifications from the Pilot Project}


\author{E. Schinnerer\altaffilmark{1} and C.L. Carilli}
\affil{NRAO, P.O. Box O, Socorro, NM 87801, U.S.A.}

\author{N.Z. Scoville}
\affil{California Institute of Technology, MS 105-24, Pasadena, CA 91125, U.S.A.}

\author{M. Bondi}
\affil{Istituto di Radioastronomia del CNR, via P. Gobetti 101, Bologna 40129, Italy}

\author{P. Ciliegi}
\affil{INAF, Osservatorio Astronomico di Bologna, via Ranzani 1, 40127 Bologna, Italy}

\author{P. Vettolani}
\affil{INAF, Osservatorio Astronomico di Bologna, via Ranzani 1, 40127 Bologna, Italy}

\author{O. Le\,Fevre}
\affil{Laboratoire d'Astrophysique de Marseille, Traverse du Siphon, 13376 Marseille Cedex 12, France }

\author{A.M. Koekemoer}
\affil{Space Telescope Science Institute, 3700 San Martin Drive, Baltimore, MD 21218, U.S.A.}

\author{F. Bertoldi}
\affil{MPI f\"ur Radioastronomie, Postfach 2024, D-53010 Bonn, Germany}

\and

\author{C.D. Impey}
\affil{Steward Observatory, University of Arizona, 933 North Cherry Avenue, Tucson, AZ 85721, U.S.A.}

\altaffiltext{1}{Jansky Postdoctoral Fellow at the National Radio Astronomy Observatory}

\begin{abstract}
We present a catalog of 246 radio sources found in the central 1
degree of the COSMOS field at 1.4\,GHz. The VLA pilot project data
have a resolution of 1.9$^{\prime\prime}$$\times$1.6$^{\prime\prime}$
and an {\sl rms} noise limit of $\sim$ 25-100\,$\mu$Jy/beam covering
0.837 deg$^2$. About 20 radio sources are clearly extended and most of
them appear to be double-lobed radio galaxies. We find evidence for a
cluster of 7 radio sources with an extent of $\sim$10$^{\prime}$
southeast of the COSMOS field center. This VLA pilot project was
undertaken to demonstrate the feasibility of wide-field mosaicking at
2'' resolution at 1.4\,GHz using the VLA in its A array
configuration. The 7-point mosaic data was used to develop the
techniques necessary for reduction and analysis. These data will
provide the initial astrometric frame for the optical (ground- and
space-based) data of the COSMOS 2\,deg$^2$ survey. In addition, it
demonstrates the feasibility of obtaining deep ({\sl rms} $\sim$ few
$\mu$Jy) radio imaging of this field at 1.4\,GHz, since the brightest
radio emission peak detected within the area covered has a flux
density of 13\,mJy/beam and no strong side-lobes from sources
surrounding the COSMOS field were detected. Comparison of the number
counts to other deep radio surveys shows that the COSMOS field is a
representative deep field in the radio domain.
\end{abstract}


\section{Introduction}

The relations between the cosmic mass distribution, environment and
galaxy properties are poorly constrained beyond a redshift of
$z\sim$0.5. Dressler et al. (1997) first demonstrated that such a
relation exists at $z\sim$0.5. The rate of galaxy evolution and the
morphological mix are thought to be strongly dependent on the local
density in the large-scale structure (LSS) but this is well
established only for the local universe (2dF and Sloan surveys;
e.g. Lewis et al. 2002, Balogh et al. 2004, G{\' o}mez et al. 2003,
Hogg et al. 2003). Substantial LSS occurs on scales up to $\sim 100$
Mpc (co-moving) (e.g. Ramella, Geller \& Huchra 1992, Gott et
al. 2003), including voids, filaments, groups and clusters which
requires wide areas and accurate spectroscopic redshifts to properly
sample it.
 
The COSMOS 2\,deg$^2$ survey (Scoville et al. 2005; {\tt
http://www.astro.caltech.edu/$\sim$cosmos}) is a pan-chromatic imaging
and spectroscopic survey of a $1.4^{\circ}\times 1.4^{\circ}$ field
designed to probe galaxy and SMBH (super-massive black hole) evolution
as a function of cosmic environment. The two major aspects of the
COSMOS survey are the HST Treasury project (P.I. Scoville), entailing
the largest ever allocation of HST telescope time -- 590 orbits for
ACS $I$ band imaging of the full field, and the extensive optical
spectroscopic survey with the VIMOS multi-object spectrograph at the
VLT with the aim of obtaining $\sim$10$^5$ spectroscopic
redshifts. The equatorial field of the COSMOS project offers the
critical advantage of allowing major observatories from both
hemispheres to join forces in this endeavor. Numerous state-of-the-art
imaging campaigns at all wavelengths (X-ray to millimeter) are
currently underway for the COSMOS field.

To study LSS it is essential to obtain high spatial resolution data
over the entire electromagnetic spectrum covering the full
2\,deg$^2$. Also, surveys of active galactic nuclei benefit from the
combination of areal coverage and depth that characterizes the COSMOS
project. To match the typical resolution for optical-NIR ground-based
data of $\sim$ 1$^{\prime\prime}$, observations with the VLA must be
conducted in the A-array which provides a resolution of about
2$^{\prime\prime}$ at 1.4\,GHz. In order to cover an area as large as
the COSMOS field mosaicking is necessary. We undertook the VLA-COSMOS
pilot project to test this approach, since no mosaicking observations
in the wide-field imaging mode with the VLA A-array at 1.4\,GHz had
been attempted yet.

The observations and data reduction are described in Section
\ref{sec:obs}. In section \ref{sec:sou} we compare and test the source 
detection algorithms used, followed by the description of the
provided catalog and the resulting source counts in section
\ref{sec:cat}. Individual sources are presented and discussed in section 
\ref{sec:gals}, and a summary is given in section \ref{sec:sum}.

\section{Observations and Data Reduction}\label{sec:obs}

\begin{figure}
\includegraphics[angle=0,scale=.75]{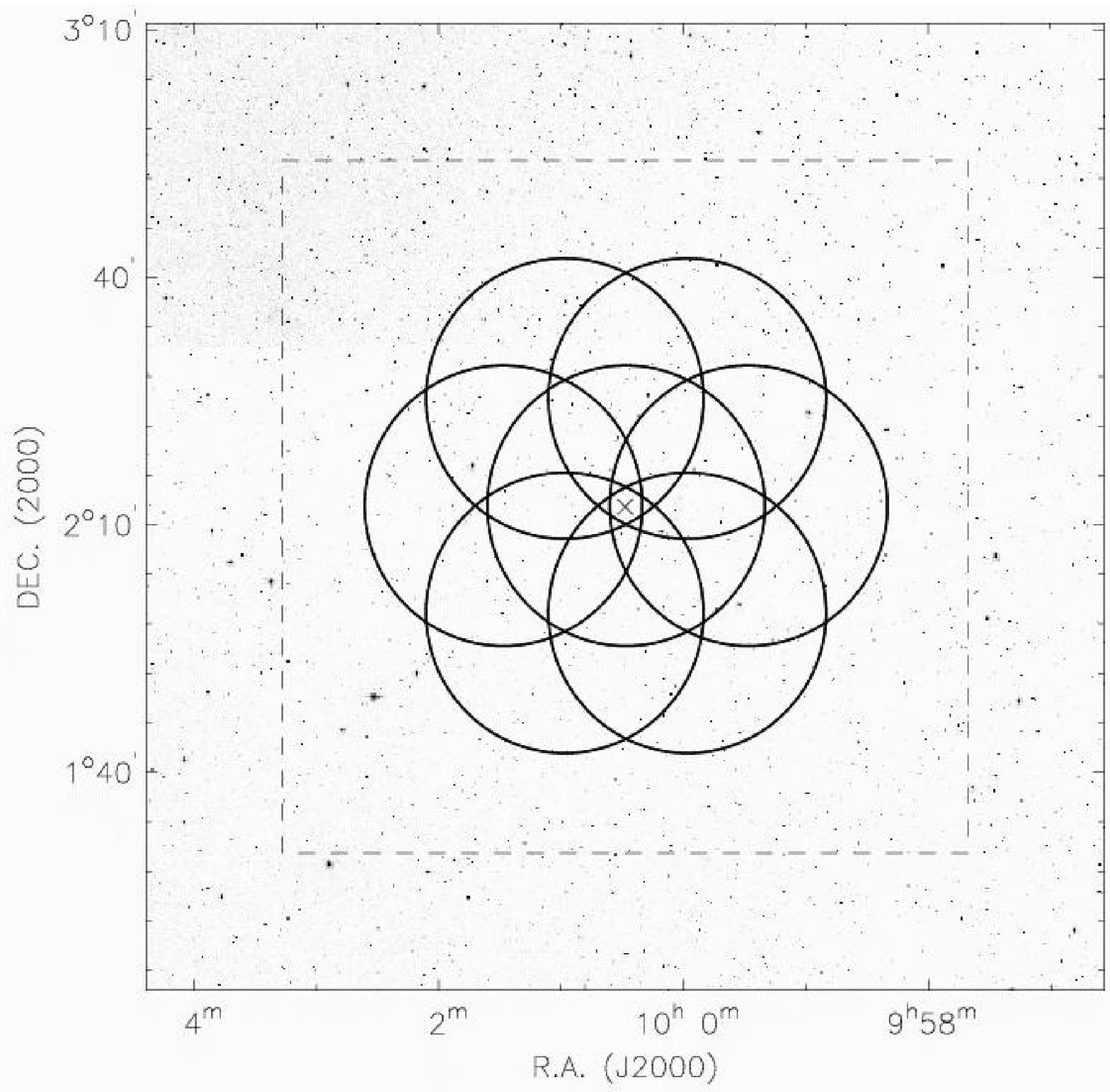}
\caption{Lay-out of the pointings of the VLA pilot project in the field
of the COSMOS survey (indicated by broken line) onto a DSS image of
the area. The cross marks the COSMOS field center. Each circle
represents the primary beam of the VLA at
1.4\,GHz.\label{fig:pattern}}
\end{figure}

The pilot VLA observations were designed to cover roughly the central
1$^{\circ}$ of the 1.4$^{\circ} \times$\,1.4$^{\circ}$ COSMOS HST
field. As described in Condon et al. (1998) a hexagonal pattern for
the pointing centers gives both an almost uniform sensitivity
distribution and a high mapping efficiency for large areas. To cover
the central degree of the COSMOS field we used 7 individual pointings
separated by 15$^{\prime}$ (the radius of the HPBW; see
Tab. \ref{tab:pos}). Our pointing separation is smaller than the
26$^{\prime}$ separation used for the 1.4\,GHz NRAO VLA Sky Survey (NVSS;
Condon et al. 1998) conducted with the VLA D-array in order to
minimize the effect of bandwidth smearing which is about
2$^{\prime\prime}$ at 15$^{\prime}$ distance from the pointing
center. The final pointing pattern is shown in Fig. \ref{fig:pattern}.

The COSMOS field was observed at 1.4\,GHz for 10hr in total on August
1st, 2003 with the NRAO Very Large Array (VLA) in its A
configuration. The standard L-Band continuum frequencies of
1.3649\,GHz and 1.4351\,GHz were used. In order to minimize the effect
of bandwidth smearing, the observations were conducted in the
multi-channel continuum mode giving two intermediate frequencies (IFs)
with 2 polarizations each. For each IF 6 channels with 3.125\,MHz were
usable for our observations resulting in a total bandwidth of
75\,MHz. The standard VLA flux calibrator 1331+305 (3C286) was
observed at the beginning and end of the observing run and served as a
flux and passband calibrator. To monitor the phase and amplitude
fluctuation we observed 1024-008 for 1.5\,min every 40\,min. 1024-008
is at a distance of about 6\,$^{\circ}$ from the COSMOS field center
with a derived flux density of 0.95 Jy at 1.4\,GHz during the time of
the observations. A typical 40min source cycle contained several
$\sim$8\,min long observations per pointing. The final on-source
integration time was about 1\,hr per field.

The data reduction was done using the NRAO AIPS package, following the
standard path for wide-field imaging and mosaicking (e.g. Condon et
al. 1998, Richards 2000). After a first atmospheric calibration using
1024-008, amplitude and phase self-calibration was employed using the
25\,-\,30 strongest (few mJy/beam) continuum sources present within
the field. In order to improve the {\sl rms} noise level, all obvious
radio frequency interference (RFI) was flagged in each channel. In
addition, we clipped all $uv$ data points above a threshold of 550
mJy/beam. A robust=1 weighting provided the optimum between {\sl rms}
sensitivity and Gaussian beam shape with a resolution of
1.9$^{\prime\prime}$$\times$1.6$^{\prime\prime}$ (PA =
-23$\deg$). Despite the equatorial field the side-lobes of the
resulting dirty beam are below the 10\% level. The final {\sl rms}
noise in each field ranged from 46 $\mu$Jy/beam (pointing 1) to 36
$\mu$Jy/beam (pointing 7) due to the very low elevation at the
beginning of the observations. For the final mosaic each pointing was
deconvolved separately using the 3-dimensional CLEAN algorithm of the
AIPS task 'IMAGR'. We CLEANed down to a residual flux level of
2.5$\sigma$.

\begin{figure}
\includegraphics[angle=0,scale=.75]{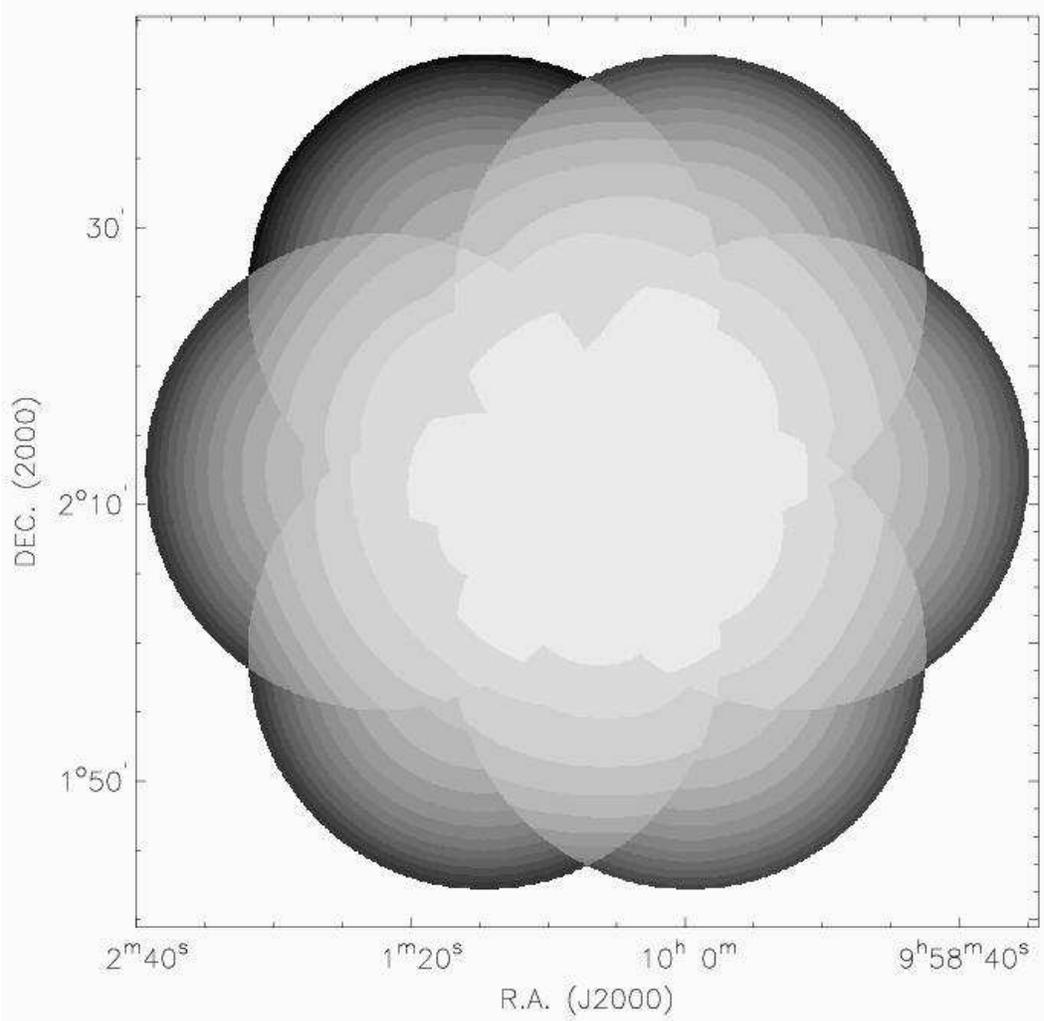}
\caption{The theoretical sensitivity map of the VLA pilot
field as constructed in AIPS showing the distribution of the {\sl rms}
noise values across the inner 1$^{\circ}$ (see section
\ref{sec:sou}). The inner 14$^{\prime}$ show a uniform {\sl rms} noise level
of $\sim$ 25$\mu$Jy/beam. Brighter colors indicate lower {\sl rms}
noise values.
\label{fig:sens}}
\end{figure}

The individual pointing images were combined using the AIPS task
'FLATN' which also takes the primary beam correction into
account. When combining images, we blanked data outside the 40\% power
radius of the primary beam. This is not quite optimal for the S/N
ratio (see Condon et al. 1998 for a discussion), however it reduces
the effect of the bandwidth smearing. (The beam broadening due to
bandwidth smearing is about 1.45 at the 40\% power radius and 2.0 at
the 10\% radius.) The variations of the {\sl rms} noise in the central
30' which have highest sensitivity (see fig. \ref{fig:sens}) are about
10\% close to the standard noise variations found for single pointings
of about 7\%.  The {\sl rms} noise in the final mosaic is
$\sim$25\,$\mu$Jy/beam for the central part and uniform out to a
radius of about 7$^{\prime}$, it increases to 100\,$\mu$Jy/beam at a
radius of about 29$^{\prime}$. The flux calibration was checked on
sources common to the NVSS (Condon et al. 1998) and individual
pointings. For each pointing we checked about 5 sources which all
showed good agreement in the flux values when taking into account the
different resolution of the NVSS and COSMOS data. The strongest radio
emission peak detected has a flux density of 13\,mJy/beam. The VLA
dynamic range due to confusing sidelobes resulting from calibration
errors is typically a few 1000. Thus this 13 mJy source will not limit
the {\sl rms} noise down to a level of a few $\mu$Jy) in the COSMOS
field.

The unique aspect of the VLA-COSMOS pilot project is its high angular
resolution of about 2$^{\prime\prime}$ compared to, for example,
6$^{\prime\prime}\,\times\,$12$^{\prime\prime}$ of the Phoenix
Deep Survey (PDS; Hopkins et al. 2003) and 6$^{\prime\prime}$ of
the Virmos Deep Field (VDF; Bondi et al. 2003). Both of these surveys
cover a comparable (VDF) or significantly larger area (PDS). The {\sl
rms} noise of the most sensitive area is better by about a factor of
1.5 (VDS) to 2 (PDS). For comparison, the full VLA-COSMOS survey is
designed to cover 2\,deg$^2$ with an {\sl rms} of $\sim$10\,$\mu$Jy in
the central 1\,deg$^2$ at a resolution of about 2$^{\prime\prime}$.

\section{Source Finding and Measurements}\label{sec:sou}

\begin{figure}
\includegraphics[angle=0,scale=.75]{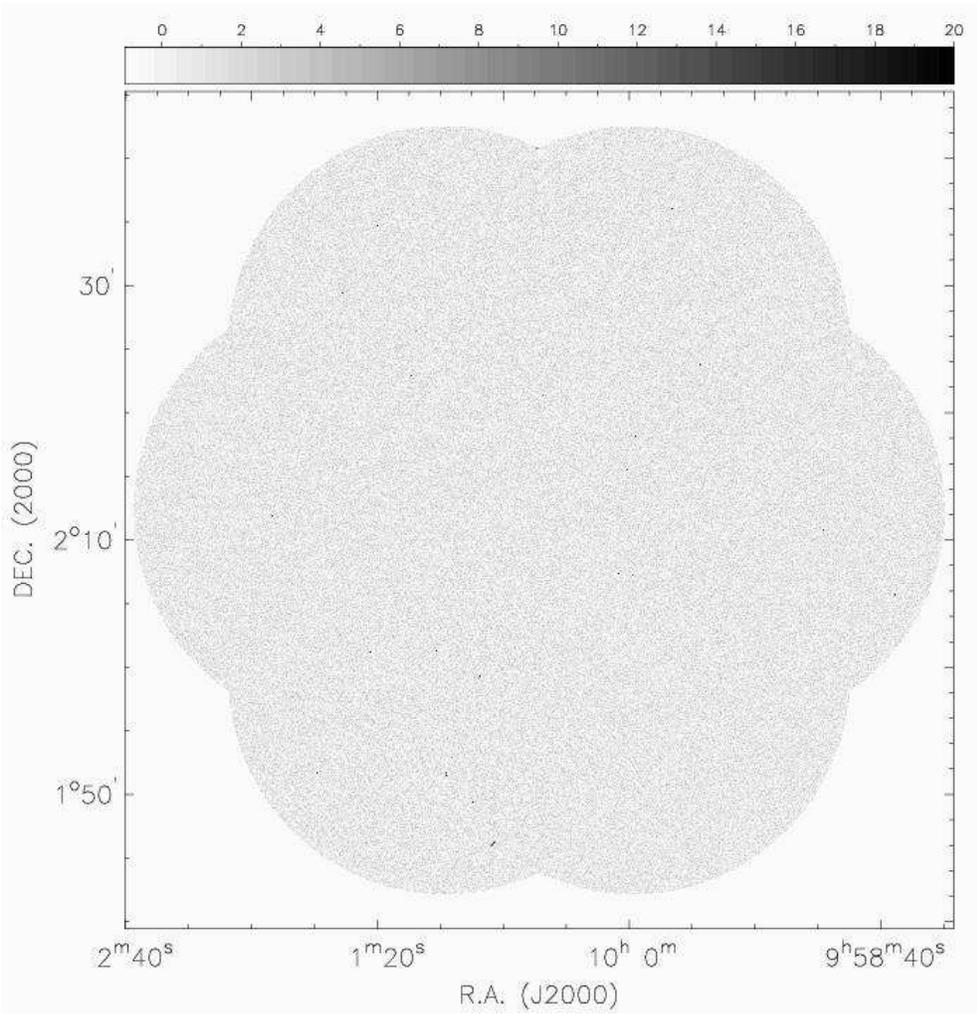}
\caption{The S/N map of the VLA COSMOS pilot
field as constructed using AIPS. The S/N map is not affected by
side-lobe effects from strong radio sources outside the mapped area.
Brighter colors indicate lower S/N values.
\label{fig:snr}}
\end{figure}

Two different source detection algorithms were thoroughly tested and
compared: 'SAD' (Search And Destroy) in the software package AIPS and
'SFIND' in the software package MIRIAD (Hopkins et al. 2002,
2003). Both tasks give parametrized results for the detected
components/sources from Gaussian fits to the data. The SAD task is
based on source detection in a simple $\sigma$-clipped image assuming
a constant {\sl rms} noise over the entire field. (Corrections for
primary beam effects can be included for single pointings only.) Thus,
in order to be able to use SAD we computed a sensitivity map
(Fig. \ref{fig:sens}) from the individual pointings using the AIPS
task 'STESS' and derived a S/N map for the entire mosaic. Then SAD was
run on the S/N map (Fig. \ref{fig:snr}) using a 5$\sigma$ detection
limit.

The task SFIND is based on the 'False Detection Rate' (FDR) algorithm
(Hopkins et al. 2002, Miller et al. 2001). The pixels for the
component/source detection are selected from a distribution of pixels
with a robustly known chance $\alpha$ (provided by the user) of being
falsely drawn from the background. We followed the approach for using
SFIND outlined by Hopkins et al. (2003). To run SFIND the mosaic was
divided into four quadrants. A box size of 100 pixels (=
47$^{\prime\prime}$) was used to determine the local {\sl rms}
noise. This size provided the best compromise between closely
following the varying sensitivity and still avoiding higher {\sl rms}
noise values close to extended radio galaxies which are mainly present
south-east of the COSMOS field center. We selected the $\alpha$
parameter to be 5\%, meaning that only 5\% of all pixels selected
should be false detections. Hopkins et al. (2002) showed that this
number is about the same for the number of false source detections. An
$\alpha$ of 5\% translates into a detection threshold of
$\sim$4.5$\sigma$ over the entire COSMOS field corresponding to
$\sim$100\,$\mu$Jy/beam in the field center which has the best
sensitivity. In addition, we did not require that for source
measurements all pixels belonging to that source lie above the
detection threshold, i.e. they had to be be contiguous and
monotonically decreasing from the peak pixel, but not necessarily to
lie above the detection threshold.

To test the effect of bandwidth smearing on the derived source size,
we made use of sources present in several individual pointings. We
compared the derived source sizes in the final mosaic to the source
sizes derived in the individual pointings. No large effect (fitted
point source size $\le$ 2$^{\prime\prime}$) of source size broadening
due to bandwidth smearing was found, as the sources which were at
large radii in one pointing were generally close to the center of
another pointing. Due to the much higher weight of the central
position, the effect of bandwidth smearing was negligible compared to
our clean beam size. Thus we conclude that bandwidth smearing is not a
dominant factor in broadening the intrinsic source sizes.

In a second test we evaluated the effects of the bandwidth smearing
and the imaging/deconvolution process on the properties (peak flux,
integrated flux, source size) of point sources in the VLA mosaic
determined by the SAD and SFIND routines. For this purpose 10 point source
models with peak values ranging from 6 to 95 $\sigma$ and sizes from
unresolved to 3$^{\prime\prime}$ were inserted in the $uv$
dataset. This $uv$ dataset was then processed in the same way as the
science data to find the inserted sources. SFIND recovered all of the
inserted sources whereas SAD did not find the 6$\sigma$ source. The
fluxes were recovered by both tasks within their uncertainties. However, both
tasks gave larger fitted source sizes for the weaker sources than the
model inputs, in particular point sources appeared resolved with
deconvolved sizes from 0.6$^{\prime\prime}$ to
1.2$^{\prime\prime}$. Thus we speculate that systematic effects in the
data (clean beam shape versus dirty beam shape, noise distribution,
etc.) have a stronger influence on the source size than bandwidth
smearing for the given observing set-up.

No systematic differences were found by determining the source sizes
using the SAD and SFIND routines or by performing the fitting 'by
hand' using the fitting tasks available in MIRIAD and AIPS. We also
compared the source positions derived by SAD and SFIND. The average
disagreement between positions in SFIND and SAD is about
0.05$^{\prime\prime}$. No systematic effect is seen. Therefore, we
conclude that our absolute astrometric uncertainty is of the order of
$\sim$0.05$^{\prime\prime}$. We decided to use the SFIND algorithm for
constructing the final source catalog, since it calculates the actual
{\sl rms} noise at the position of the source and it seemed more
sensitive to the weakest sources.

\section{Catalog and Counts}\label{sec:cat}

\begin{figure}
\includegraphics[angle=-90,scale=.65]{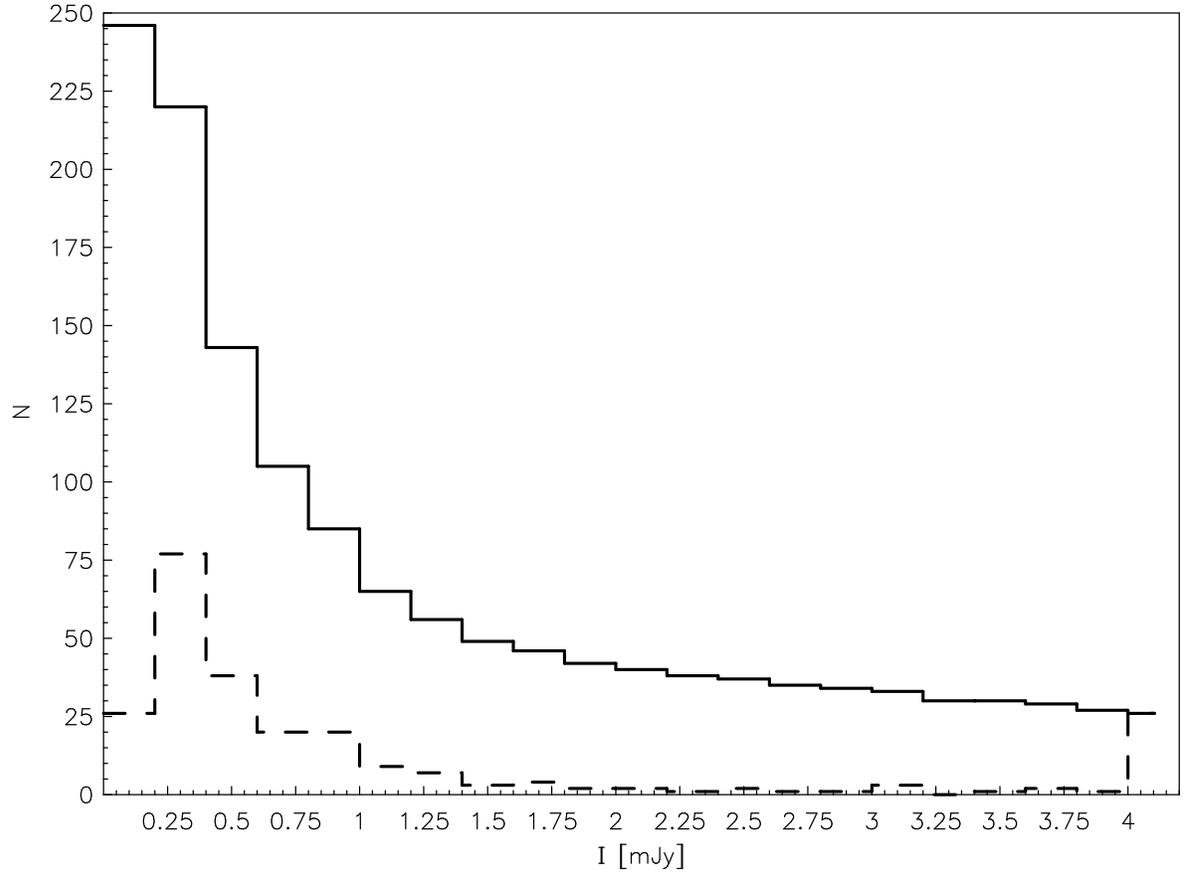}
\caption{Distribution of the observed integrated flux densities 
(broken line). The drop of sources in the bin of our 4.5$\sigma$ limit
(I $<$ 0.2\,mJy $\approx$ 8$\sigma$)
indicates that the completeness is fairly low at the faintest flux
limit in the COSMOS catalog. The cumulative numbers are shown as well
(solid line). All sources with fluxes $>$ 4\,mJy are added together in
the last bin at 4.1\,mJy.
\label{fig:flux}}
\end{figure}

The final catalog (see section \ref{sec:cat} and Tab. \ref{tab:list})
was constructed from an initial list of 305 components found by the
routine SFIND for a False Detection Rate of 5\% (see section
\ref{sec:sou}). (Note that about 5\% of these components should be
false detections, as already mentioned in section
\ref{sec:sou}.) We identified by eye 20 sources (or groups) which were
fitted by multiple (sometimes even overlapping) Gaussian
components. These 20 groups are displayed in Fig. \ref{fig:radio} -
\ref{fig:radio5}), Tab. \ref{tab:radio} lists the number of Gaussian
components found for each of these sources. For these groups, the
position and flux density of the emission peak of the entire group of
components replaced the entries of the individual components in the
final source catalog. The integrated flux of a group was derived using
the AIPS task TVSTAT which allowed us to integrate over the irregular
area of these sources giving a non-parametric result. One obvious
misidentification on a side-lobe was rejected as well. The final
catalog contains a total of 246 sources. The distribution of sources
with respect to the integrated flux density shows that the
completeness of the catalog at the faintest flux levels ($\le$
250$\mu$Jy) is fairly low (Fig. \ref{fig:flux}).

\subsection{Uncertainty Estimate}

The uncertainties on the integrated and peak flux densities determined by a
Gaussian fit are in general smaller than the true uncertainties. The
error propagation equations by Condon (1997) which assume that
Gaussian random noise is dominating the uncertainties in the data (Condon
1997) can be used to estimate the true uncertainties. We followed the
approach given by Hopkins et al. (2003) based on the assumption that
the relative uncertainty $\frac{\sigma_I}{I}$ in the integrated flux density 
is due to uncertainties $\mu_{data}$ in the data and uncertainties
$\mu_{fit}$ in the Gaussian fit (their equation 1):

\begin{equation}
\frac{\sigma_I}{I} = \sqrt{\left (\frac{\mu_{data}}{I}\right )^2 +
\left (\frac{\mu_{fit}}{I}\right )^2}\,\,.
\label{equ1}
\end{equation}

The relative uncertainty $\frac{\mu_{data}}{I}$ in the data is given by
Windhorst, van Heerde \& Katgert (1984):

\begin{equation}
\frac{\mu_{data}}{I} = \sqrt{\left (\frac{{\sl rms}}{S}\right )^2 + 0.01^2}
\label{equ2}
\end{equation}

where $S$ and {\sl rms} are peak flux density and noise at the
position of the source. The constant term represents the relative
uncertainties in the absolute flux calibration and due to pointing
errors of the individual telescopes which are together of the order of
1\%.

Equation (42) of Condon (1997) gives a relation between the relative
uncertainty from the fitting and the {\sl rms} noise in the image
which is correlated over the synthesized beam area. Following Hopkins
et al. (2003; see their equation 3), we use the product of the major
and minor axis of the full width half maximum of the beam ($\theta_B
\theta_b$) and the measured source size ($\theta_M \theta_m$). The 
fitting of the peak flux $S$, major axis $\theta_M$ and minor axis
$\theta_m$ has the relative uncertainties $\frac{\mu_S}{S}$,
$\frac{\mu_M}{\theta_M}$ and $\frac{\mu_m}{\theta_m}$,
respectively. These can be approximated by $\left
(\frac{\mu_X}{X}\right )^2 \approx \frac{2}{\rho_X^2}$ (equation 21 by
Condon 1997) with $X = S, M$ or $m$. Thus equation (42) of Condon
(1997) can be written as:

\begin{equation}
\frac{\mu_{fit}}{I} = \sqrt{\frac{2}{\rho_S^2} +
\left(\frac{\theta_B \theta_b}{\theta_M \theta_m}\right)
\left(\frac{2}{\rho_M^2} + \frac{2}{\rho_m^2}\right)}\,\,.
\label{equ3}
\end{equation}

The signal-to-noise ratios (S/N) of the fit, $\rho_S$, $\rho_M$ and $\rho_m$,
are parameter dependent (see equation 41 of Condon 1997). Following
Hopkins et al. (2003) and Condon (1997), they can be calculated as follows

\begin{equation}
\rho_X^2 = \frac{\theta_M \theta_m}{4 \theta_B \theta_b} 
\left [1 + \left (\frac{\theta_B}{\theta_M}\right )\right ]^{\alpha}
\left [1 + \left (\frac{\theta_b}{\theta_m}\right )\right ]^{\beta}
\left (\frac{S}{{\sl rms}}\right )^2
\label{equ4}
\end{equation}

with $\alpha=\beta=1.5$ for $\rho_S^2$, $\alpha=2.5$ and $\beta=0.5$
for $\rho_M^2$, and $\alpha=0.5$ and $\beta=2.5$ for $\rho_m^2$.

\subsection{Derived Source Sizes}

The deconvolved source sizes were derived from the fitted sizes given
by SFIND and the size of the clean beam. Since we did not check the
PSF during the size fit with SFIND, some fitted source sizes are
smaller than the clean beam. These sizes were excluded from the size
deconvolution. However, the effect on the integrated flux is minimal
compared to enforcing a minimal source size of the clean beam.

\begin{figure}
\includegraphics[angle=-90,scale=.65]{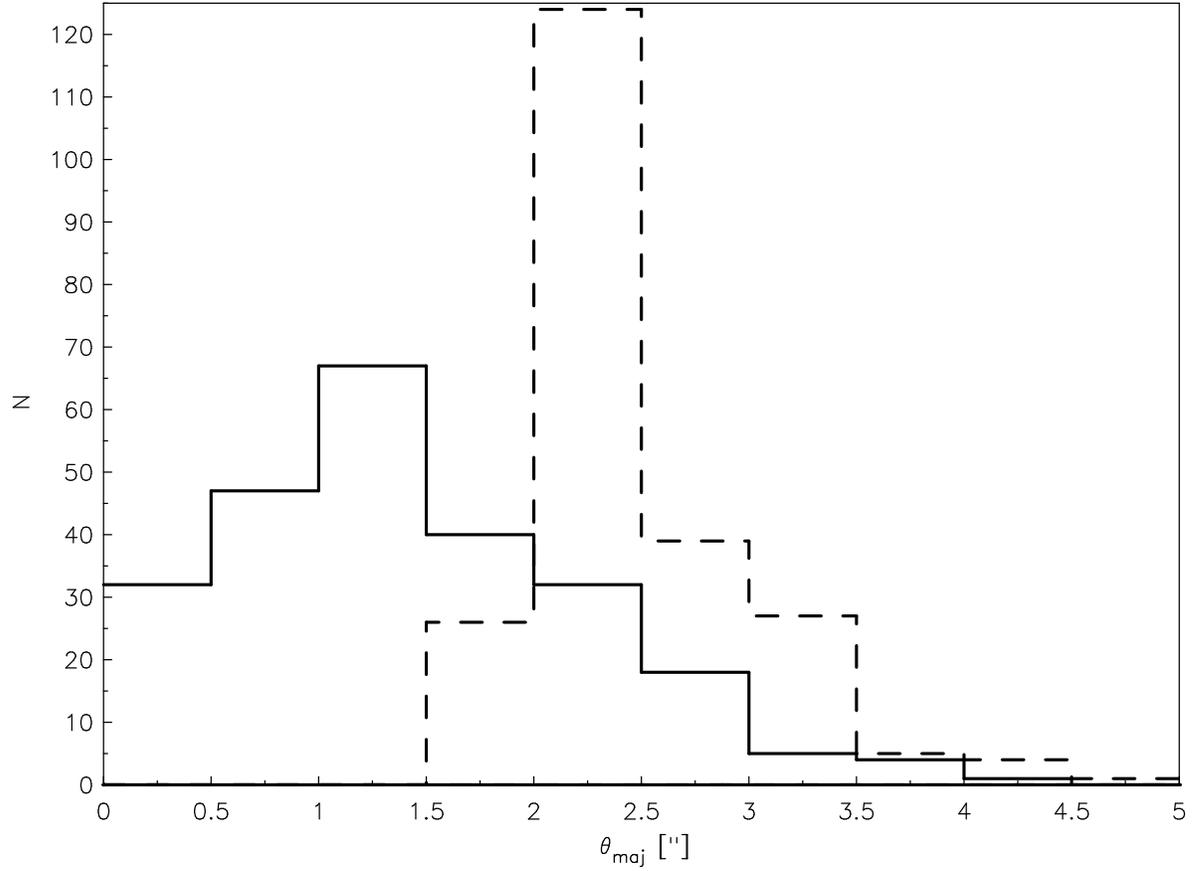}
\caption{Distribution of the fitted (broken line) and deconvolved 
(solid line) source sizes. The large number of sources with fitted
source sizes around 2$^{\prime\prime}$ (equal to about
1.25$^{\prime\prime}$ deconvolved source sizes) suggests that only
fitted sizes $\ge$ 2.5$^{\prime\prime}$ can be trusted (see text for
details).
\label{fig:size}}
\end{figure}

A significant fraction of our sources appear resolved (see
Fig. \ref{fig:size}), however, as discussed in section \ref{sec:sou}
the derived values for the faint and small sources might not be
correct. From Fig. \ref{fig:size}, we estimate that fitted source
sizes of $\le$ 2.5$^{\prime\prime}$ are not reliable. This is in
agreement with our modeling. Note that Richards (2000) concluded after
thorough testing and modeling that only these sources close to the
detection limit with sizes of $\ge$ 2.7$^{\prime\prime}$ could be
reliably resolved in the $\sim$2$^{\prime\prime}$ VLA A-array data of
the HDF-N field. Therefore the size distribution for weak radio
sources (with fitted sizes smaller than about 2.5$^{\prime\prime}$) in
our VLA data remains largely unknown.

\subsection{Description of the Catalog}

The final catalog is presented in Tab. \ref{tab:list}. Sources which
were fitted by multiple Gaussian components are presented in detail in
Section \ref{sec:gals}, in Tab. \ref{tab:list} we list the position of
the emission peak as well as the derived integrated flux for these
sources. (Note that the absolute astrometric uncertainty is of the
order of 0.05$^{\prime\prime}$; see section \ref{sec:sou}.) All 246
radio sources are listed in right ascension order in
Tab. \ref{tab:list} with the following columns:
\\
\\
Column(1): Right ascension (J2000.0) and its {\sl rms} uncertainty
\\
Column(2): Declination (J2000.0) and its {\sl rms} uncertainty
\\
Column(3): Peak flux density and its {\sl rms} uncertainty
\\
Column(4): Integrated flux density and its {\sl rms} uncertainty
\\
Column(5): {\sl rms} measured by SFIND at the position of the radio
source
\\
Column(6): Fitted source size -- major axis $\theta_{M,fit}$
\\ 
Column(7): Fitted source size -- minor axis $\theta_{m,fit}$ 
\\
Column(8): Fitted source size -- $PA_{fit}$
\\
Column(9): Deconvolved source size -- major axis $\theta_{M,dec}$
\\
Column(10): Deconvolved source size -- minor axis $\theta_{m,dec}$ 
\\
Column(11): Deconvolved source size -- $PA_{dec}$
\\
Column(12): Flag for sources with multiple components which are presented in
Section \ref{sec:gals}.

\subsection{Number Counts}

To derive the number counts, we divided our source catalog into six
bins with about 40 sources each to provide reasonable statistics. This
is intended to test whether the radio source counts in the COSMOS
field are similar to the ones obtained in other deep
radio fields.

In order to derive the correct source counts at a given flux density,
we need to correct for the fact that the {\sl rms} noise across the field is
varying (weighting correction) and that weaker extended source will be
missed as their peak flux density is below our detection threshold
while their integrated flux density might still be above (resolution
effect). The relation between the measured source counts $N$ and the
corrected source counts $N_{eff}$ is:

\begin{equation}
N_{eff} = N \times w \times r\,\,.
\label{equ5}
\end{equation}

\begin{figure}
\includegraphics[angle=-90,scale=.65]{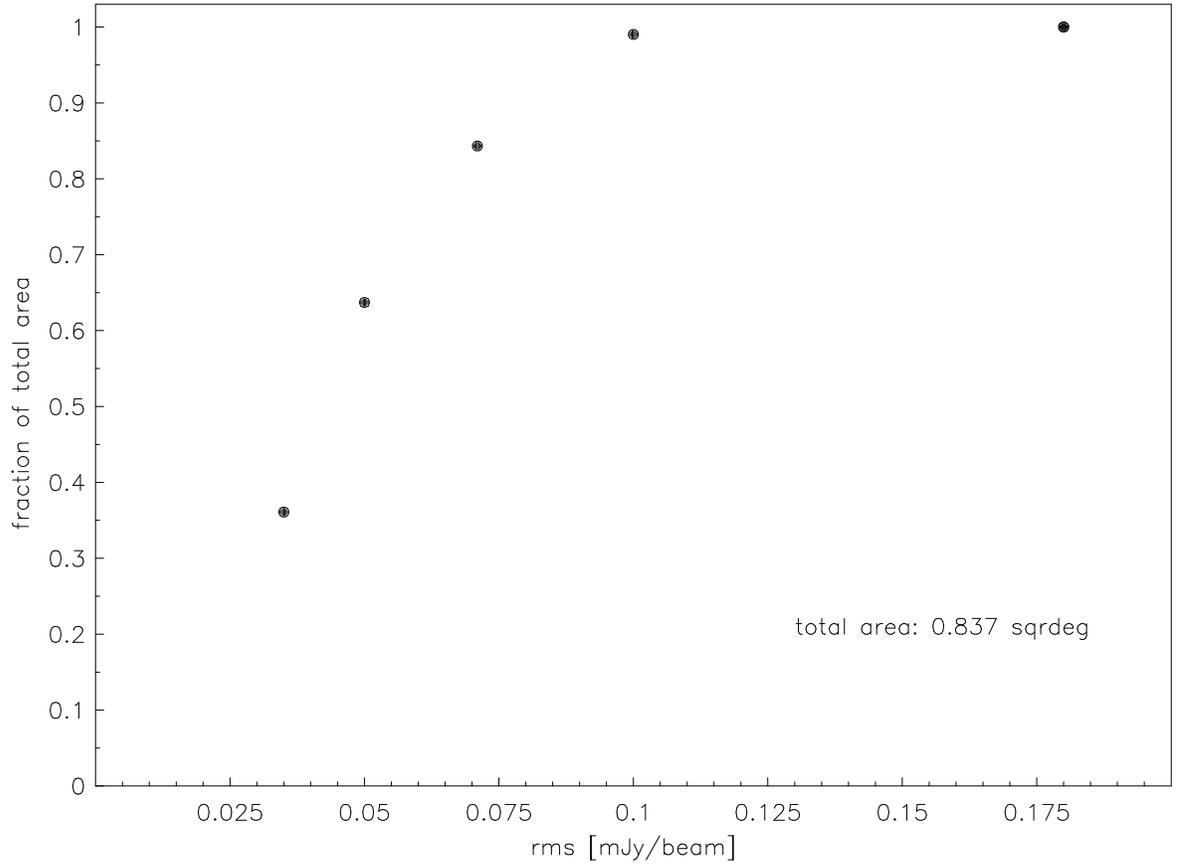}
\caption{The {\sl rms} noise level versus the (cumulative) fraction of the
area covered. The full area covered is 0.837 deg$^2$.
\label{fig:area}}
\end{figure}

The weighting correction $w$ simply depends on the effective area $D$ for
source detection with respect to the total area $T$ covered by the
survey (see Fig. \ref{fig:area}):

\begin{equation}
w = T / D
\label{equ6}
\end{equation}

Windhorst et al. (1990) gives the following relation for the
distribution of the source sizes $h(\Psi)$ at a given flux density (see also
Hopkins et al. 1998):

\begin{equation}
h(\Psi) = e^{-ln\left [2 \left (\frac{\Psi}{\Psi_{med}}\right )^{0.62}\right ]}
\hspace{15mm} {\rm with} \hspace{5mm}
\Psi_{med} = 2.0'' \times S_{1.4GHz}^{0.30}
\label{equ7}
\end{equation}

where $\Psi_{med}$ is the medium source size at a given flux density
$S_{1.4GHz}[mJy]$ at 1.4\,GHz. Using $\Psi=\Psi_{max}$ for the maximal
detectable structure at a given flux density allows us to derive the
correction factor $r$ for large sources missed in the data:

\begin{equation}
r = \frac{1}{1 - h(\Psi_{max})}\,\,.
\label{equ8}
\end{equation}

Windhorst et al. (1993) derived an average source size of about
2.0$^{\prime\prime}$ for sub-mJy sources. Richards (2000) and Bondi et
al. (2003) find similar values, however Bondi et al. (2003) found that
the Windhorst et al. (1990) relation over-predicts the number of
sources with large angular sizes through comparison to their data on
the VDF field. To allow for easy comparison, we use the Windhorst et
al. (1990) relation with an average source size of
2.0$^{\prime\prime}$ (see Tab. \ref{tab:counts}).

\begin{figure}
\includegraphics[angle=-90,scale=.65]{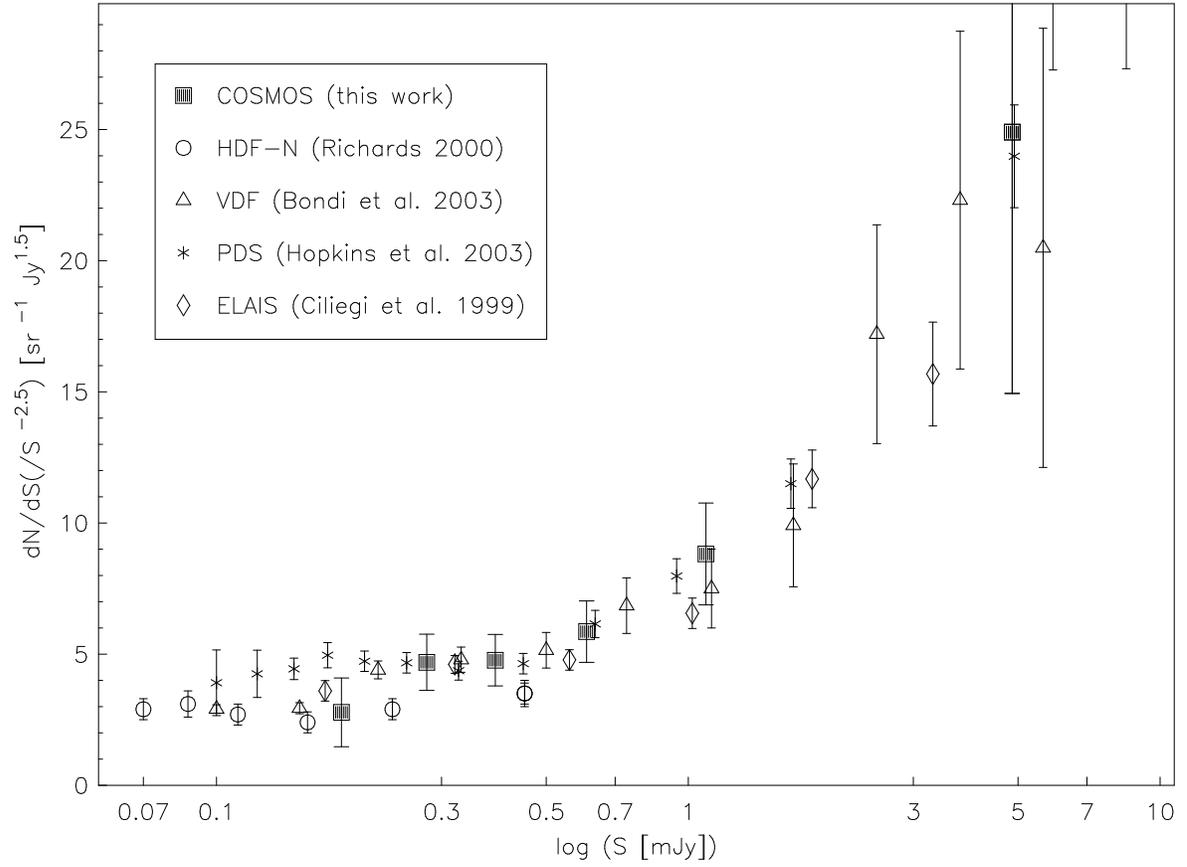}
\caption{The Euclidean normalized source counts dN/dS(/S$^{-2.5}$) for the
VLA-COSMOS field at 1.4\,GHz. The counts for the COSMOS field are in
agreement with the count numbers found for other radio deep
fields. 
\label{fig:counts}}
\end{figure}

We present the Euclidean normalized differential source counts
$\frac{dN}{dS}(/S^{2.5})$ for the COSMOS field in
Fig. \ref{fig:counts}. The comparison with results of similar deep
field surveys (HDF-N: Richards et al. 2000; VDF: Bondi et al. 2003,
PDS: Hopkins et al. 2003, ELAIS: Ciliegi et al. 1999) shows that the
COSMOS field is representative.

\section{The central 9$^{\prime}$$\times$9$^{\prime}$ HST ACS Field and Individual Radio Sources}\label{sec:gals}

\begin{figure}
\includegraphics[angle=0,scale=.75]{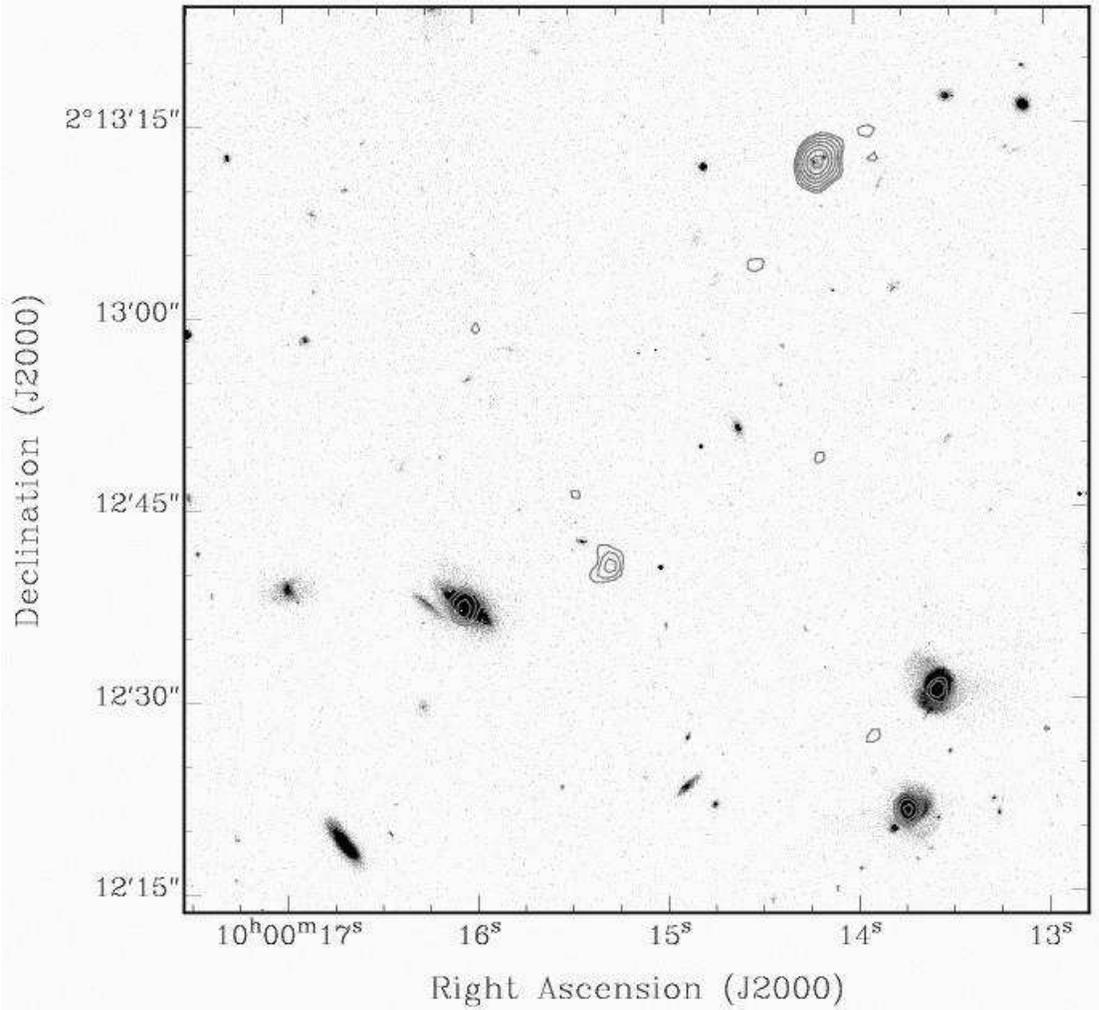}
\caption{Comparison between an apparent cluster of sources to the
optical data. VLA-COSMOS 1.4\,GHz data overlaid in contours onto a
sub-image of the HST ACS $I$ band data from the inner
9$^{\prime}$$\times$9$^{\prime}$ field. About 75\% of the radio
sources in the inner 9$^{\prime}$$\times$9$^{\prime}$ area have
optical counterparts in the HST data. The optical counterparts show a
variety in host galaxy properties demonstrating the potential in using
morphology information. The contours are in steps of
1$\sigma$=25$\mu$Jy/beam starting at 3$\sigma$.
\label{fig:hst}}
\end{figure}

We have detected 12 radio sources within the area covered by the
initial central 9$^{\prime}$$\times$9$^{\prime}$ of the HST ACS $I$
and $g$ band data. (Note that the magnitude limit for the HST ACS $I$
band survey is $A_{AB} \sim 27$\,mag for a 5\,$\sigma$ point source.) 
We find 9 optical counterparts suggesting that about 75\% of all our
radio sources will have optical counterparts in the HST images. The
resolved optical sources range from knotty spiral galaxies, one
apparently interacting galaxy pair to early type galaxies
(Fig. \ref{fig:hst}). This clearly demonstrates that the combination
of the HST ACS data and the VLA radio data (together with ground-based
optical imaging and spectroscopic data as well as XMM X-ray data) will
allow classification of the origin of the radio emission (star
formation vs. AGN) based on host galaxy properties.
 
\begin{figure}
\includegraphics[angle=-90,scale=.65]{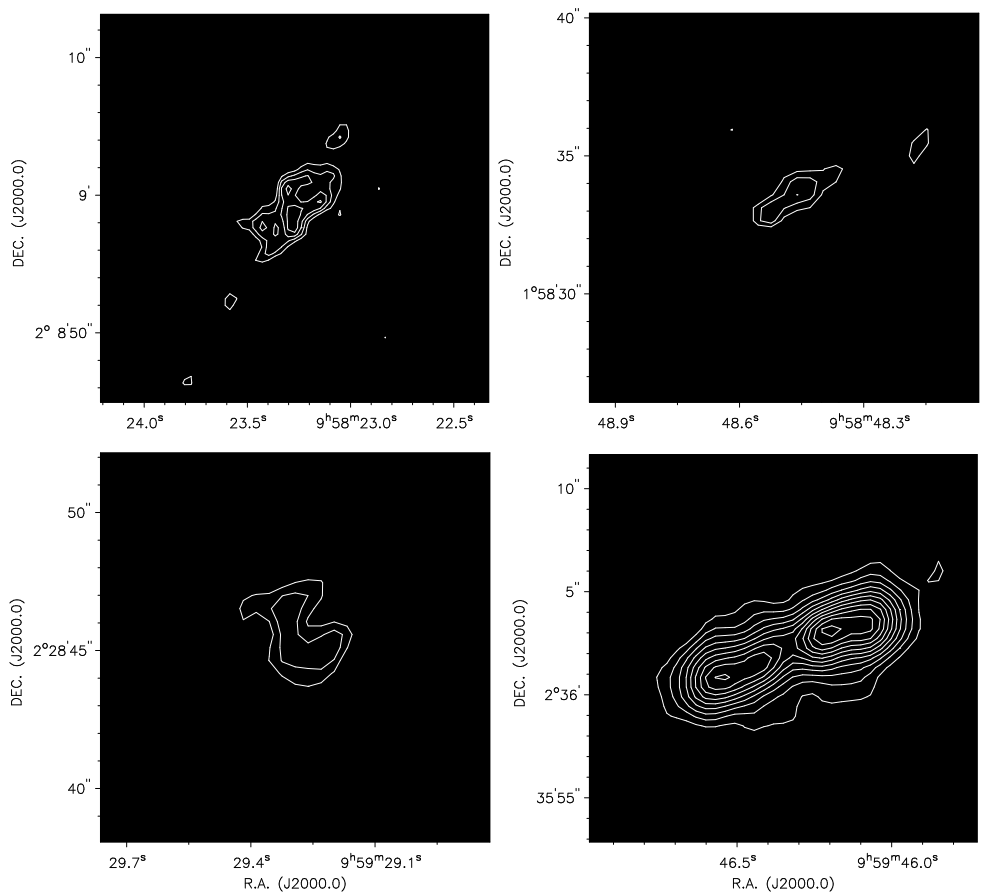}
\caption{All radio sources fitted by multiple Gaussian components and
identified as a radio group (see Tab. \ref{tab:radio}). The contours
are in steps of 1$\sigma$ starting at 3$\sigma$. 
(With 1$\sigma$ equals to 
83 $\mu$Jy/beam ({\it top left}),
70 $\mu$Jy/beam ({\it top right}),
40 $\mu$Jy/beam ({\it bottom left}), and
56 $\mu$Jy/beam ({\it bottom right}).)
\label{fig:radio}}
\end{figure}

\begin{figure}
\includegraphics[angle=-90,scale=.65]{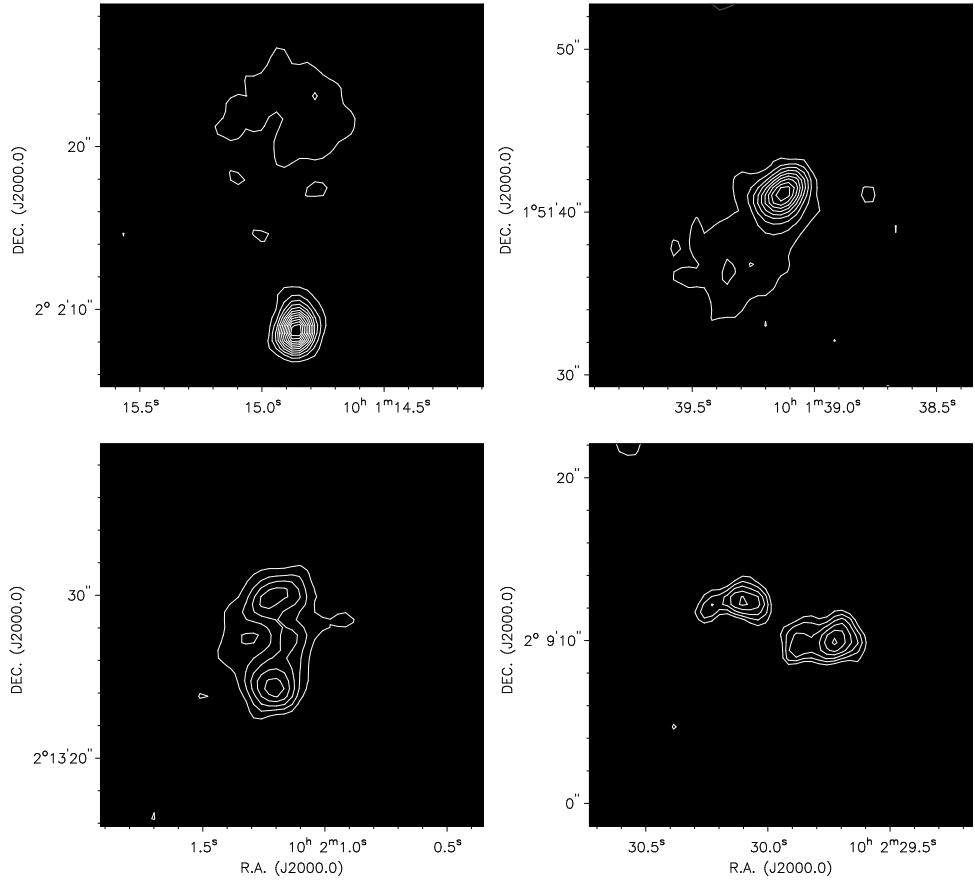}
\caption{Same as Fig. \ref{fig:radio}.
(With 1$\sigma$ equals to 
32 $\mu$Jy/beam ({\it top left}),
68 $\mu$Jy/beam ({\it top right}),
49 $\mu$Jy/beam ({\it bottom left}), and
84 $\mu$Jy/beam ({\it bottom right}).)
\label{fig:radio5}}
\end{figure}

We identified 20 radio sources which were fit by more than one
Gaussian component present in the pilot data-set (Fig. \ref{fig:radio}
to \ref{fig:radio5}). These radio groups are listed in
Tab. \ref{tab:radio}. Most of them are very likely FRII radio galaxies
exhibiting the typical double-lobe structure. Not randomly
distributed, an apparent cluster of 6 radio galaxies is seen to the
south-east of the COSMOS field center with an extent of
$\sim$10$^{\prime}$. The five double-lobed radio galaxies (ID\# 134,
138, 159, 181 and 182 in Tab. \ref{tab:radio}) and the 1 single-lobed
radio galaxy (ID\# 141 in Tab. \ref{tab:radio}) of this apparent
cluster are shown in Fig. \ref{fig:radio} to \ref{fig:radio5}.

\section{Summary and Conclusions}\label{sec:sum}

We have presented the first wide-field imaging mosaic obtained with
the VLA in A configuration at 1.4\,GHz. These data cover the inner 1
degree of the COSMOS field down to an {\sl rms} noise level of
$\sim$25\,$\mu$Jy/beam in the inner region. We present a radio source
catalog containing 246 entries above the 4.5$\sigma$ limit. About 20
sources are well-resolved into multiple components likely being
double-lobed radio galaxies. An apparent cluster of 6 radio galaxies
is found to the southeast of the COSMOS field containing 5
double-lobed radio galaxies. No strong radio emission peak was found
in the field which could prevent further deep (down to the few $\mu$Jy
level) radio imaging of the COSMOS field. The radio number counts are
consistent with those derived for other fields such as the Phoenix
deep field and the VIRMOS deep field showing that the COSMOS field is
a representative field in the radio domain.

Comparison between the source finding algorithms SAD in AIPS and SFIND
in MIRIAD shows similar results for the fitting of the source
properties, however, SFIND was more sensitive to sources close to our
detection limits. We found that only fitted source sizes
$\ge$2.5$^{\prime\prime}$ are reliable in our data. The full
VLA-COSMOS source catalog as well as the image are available in
electronic form from the COSMOS archive at IPAC/IRSA ({\tt
http://www.irsa.ipac.caltech.edu/data/COSMOS} starting mid-Aug
2004).

In the future, the COSMOS archive will also contain (photometric and
spectroscopic) redshifts, optical morphologies, and flux densities
from the X-ray to millimeter wavelengths.

\acknowledgments

Special thanks to A. Hopkins for his advice on the use of SFIND. We
would also like to thank F. Owen and E. Greisen for helpful discussion
on the data processing. ES and CC acknowledge support from NASA grant
HST-GO-09822.31-A. 


\clearpage

\begin{figure}
\includegraphics[angle=-90,scale=.65]{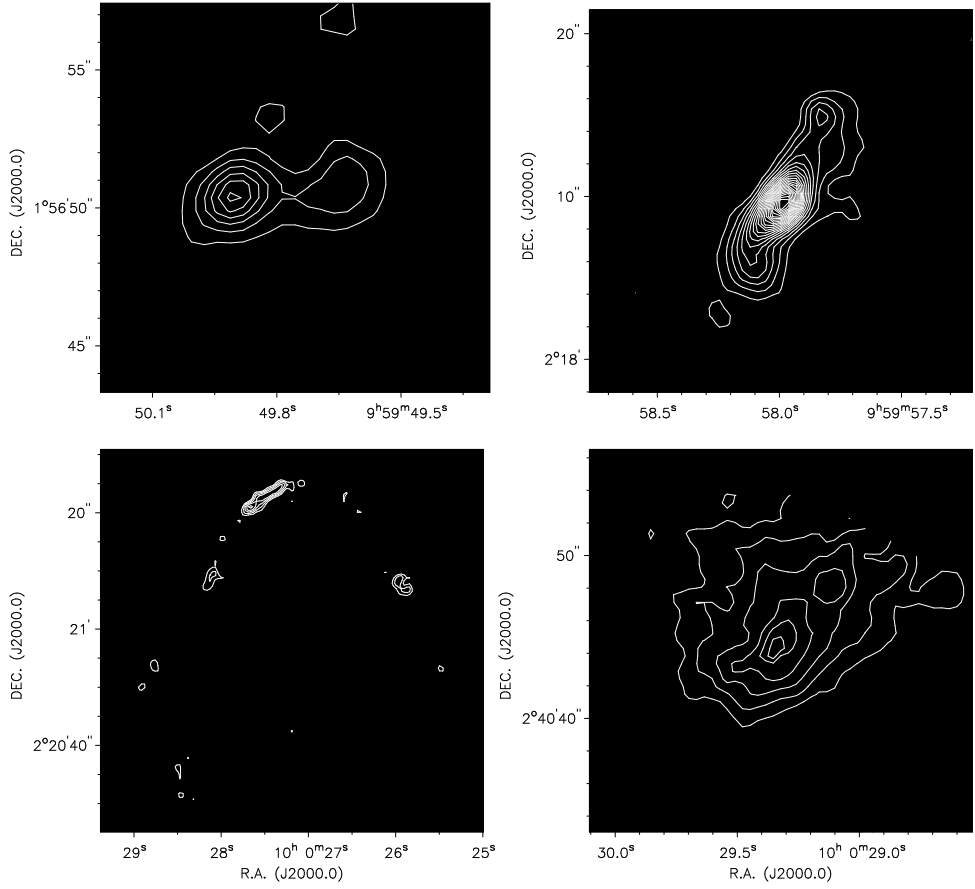}
\caption{Same as Fig. \ref{fig:radio}. The source shown in the {\it
bottom right} panel is very likely the southern lobe of a radio galaxy
located outside our area analyzed.
(With 1$\sigma$ equals to 
32 $\mu$Jy/beam ({\it top left}),
28 $\mu$Jy/beam ({\it top right}),
34 $\mu$Jy/beam ({\it bottom left}), and
100 $\mu$Jy/beam ({\it bottom right}).)
}
\end{figure}

\begin{figure}
\includegraphics[angle=-90,scale=.65]{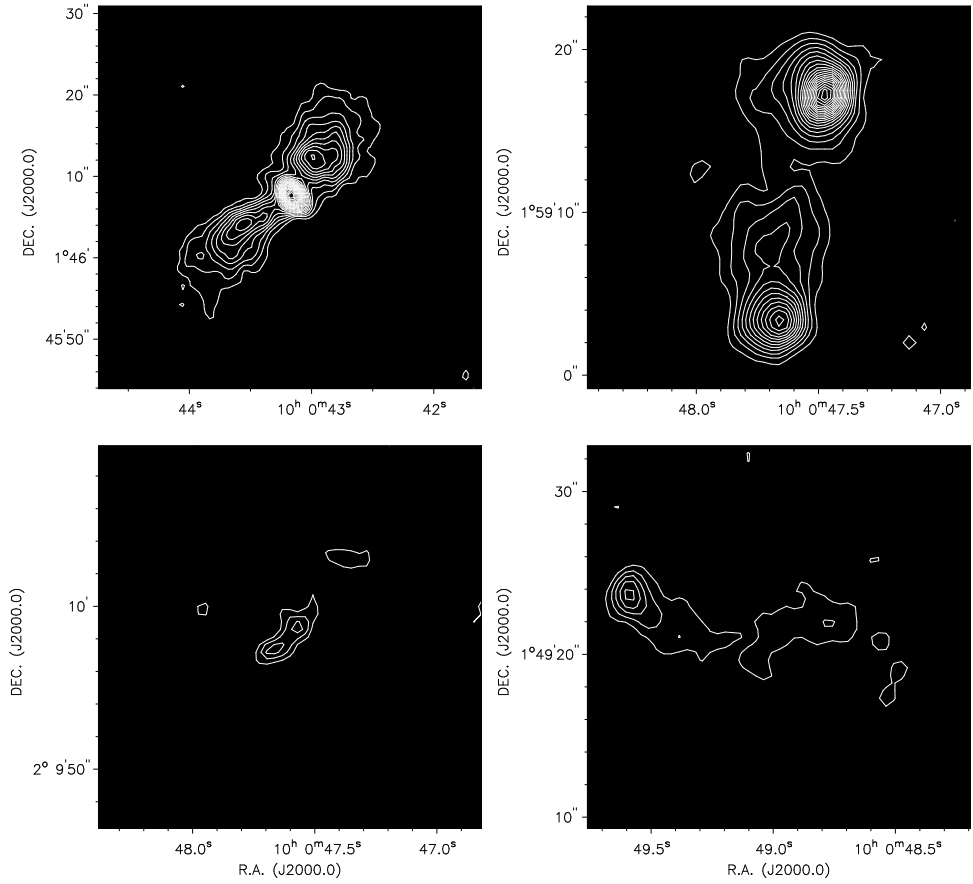}
\caption{Same as Fig. \ref{fig:radio}.
(With 1$\sigma$ equals to 
70 $\mu$Jy/beam ({\it top left}),
33 $\mu$Jy/beam ({\it top right}),
25 $\mu$Jy/beam ({\it bottom left}), and
46 $\mu$Jy/beam ({\it bottom right}).)
}
\end{figure}

\begin{figure}
\includegraphics[angle=-90,scale=.65]{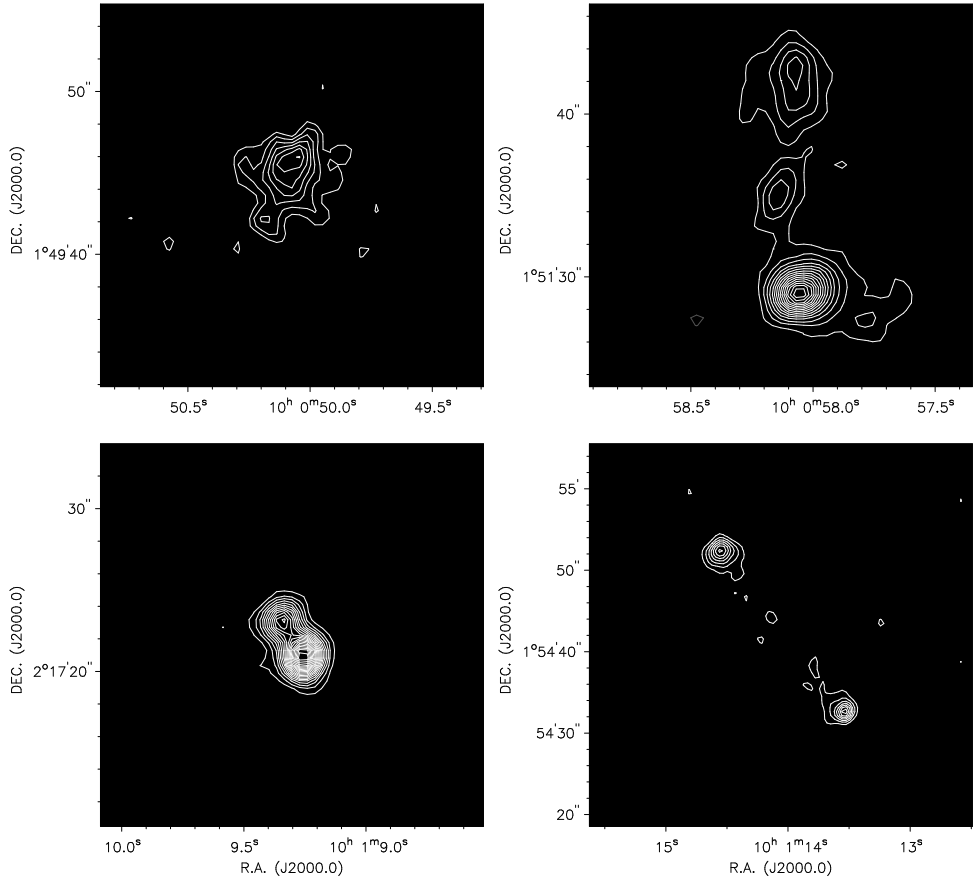}
\caption{Same as Fig. \ref{fig:radio}.
(With 1$\sigma$ equals to 
43 $\mu$Jy/beam ({\it top left}),
45 $\mu$Jy/beam ({\it top right}),
30 $\mu$Jy/beam ({\it bottom left}), and
45 $\mu$Jy/beam ({\it bottom right}).)
}
\end{figure}

\clearpage

\begin{deluxetable}{crr}
\tablecaption{VLA Pointing Centers\label{tab:pos}}
\tablewidth{0pt}
\tablehead{
\colhead{Pointing \#} & 
\colhead{R.A. (J2000)} & 
\colhead{DEC (J2000)}}
\startdata
1 &  10:00:58.62  &  +02:25:20.42  \\
2 &  09:59:58.58  &  +02:25:20.42  \\
3 &  10:01:28.64  &  +02:12:21.00   \\
4\tablenotemark{a} &  10:00:28.60  &  +02:12:21.00   \\
5 &  09:59:28.56  &  +02:12:21.00   \\
6 &  10:00:58.62  &  +01:59:21.58  \\
7 &  09:59:58.58  &  +01:59:21.58  \\
\enddata
\tablenotetext{a}{COSMOS field center}
\tablecomments{Pointing centers for the VLA pilot project at 1.4\,GHz.}
\end{deluxetable}

\clearpage

\begin{deluxetable}{llllcrrrrrrc}
 \tabletypesize{\scriptsize}
\rotate
\tablecaption{1.4\,GHz Source Catalog of the COSMOS VLA Pilot Project\label{tab:list}}
\tablewidth{0pt}
\tablehead{
\colhead{R.A.} & \colhead{Dec.} & \colhead{S} & \colhead{I} &
\colhead{{\sl rms}} & \colhead{$\theta_{M,fit}$} &
\colhead{$\theta_{m,fit}$} &\colhead{$PA_{fit}$} & \colhead{$\theta_{M,dec}$} &
\colhead{$\theta_{m,dec}$} &\colhead{$PA_{dec}$} & \colhead{Flag}
\\
\colhead{(J2000.0)} & \colhead{(J2000.0)} & \colhead{[mJy/beam]} &
\colhead{[mJy]} & \colhead{[mJy/beam]} & \colhead{[$^{\prime\prime}$]} &
\colhead{[$^{\prime\prime}$]} &\colhead{[$^o$]} &\colhead{[$^{\prime\prime}$]} & \colhead{[$^{\prime\prime}$]} & 
\colhead{[$^o$]} & }
\startdata
09:58:23.271$\pm$ $\dots$ & 2:08:57.63$\pm$   $\dots$  &  0.564$\pm$ $\dots$ &   5.765$\pm$ $\dots$ &  0.083 &  0.0 &  0.0 &   0.0 &  0.0 &  0.0 &   0.0 & y \\
09:58:28.708$\pm$   0.010 & 2:11:00.19$\pm$      0.00  &  0.530$\pm$  0.117  &   0.825$\pm$  0.227  &  0.072 &  2.9 &  1.7 &  88.4 &  2.4 &  0.0 &  85.2 &   \\
09:58:29.067$\pm$   0.006 & 2:05:31.20$\pm$      0.01  &  0.683$\pm$  0.147  &   1.779$\pm$  0.441  &  0.089 &  3.2 &  2.5 &  45.3 &  2.8 &  1.7 &  49.6 &   \\
09:58:34.874$\pm$   0.012 & 2:03:12.49$\pm$      0.01  &  0.561$\pm$  0.141  &   1.228$\pm$  0.367  &  0.088 &  3.2 &  2.1 &  62.1 &  2.8 &  0.9 &  62.8 &   \\
09:58:35.454$\pm$   0.001 & 2:05:43.65$\pm$      0.00  &  6.477$\pm$  0.657  &  10.701$\pm$  1.095  &  0.074 &  2.6 &  2.0 &  64.1 &  2.0 &  0.6 &  64.9 &   \\
09:58:35.760$\pm$   0.014 & 2:05:46.07$\pm$      0.01  &  0.385$\pm$  0.097  &   0.318$\pm$  0.115  &  0.066 &  2.3 &  1.1 & -46.3 &  1.4 &  0.0 & 127.3 &   \\
09:58:35.916$\pm$   0.003 & 2:12:33.13$\pm$      0.00  &  0.350$\pm$  0.089  &   0.285$\pm$  0.103  &  0.060 &  1.6 &  1.5 & -13.9 &  0.0 &  0.0 &   0.0 &   \\
09:58:42.056$\pm$   0.006 & 2:15:10.94$\pm$      0.00  &  0.391$\pm$  0.090  &   0.618$\pm$  0.177  &  0.055 &  2.5 &  1.9 & -66.8 &  1.8 &  0.7 & 102.2 &   \\
09:58:44.728$\pm$   0.004 & 2:02:49.57$\pm$      0.01  &  0.553$\pm$  0.120  &   1.086$\pm$  0.283  &  0.072 &  2.8 &  2.1 &   3.4 &  2.1 &  1.3 &  11.7 &   \\
09:58:45.264$\pm$   0.005 & 2:15:40.90$\pm$      0.01  &  0.453$\pm$  0.090  &   0.815$\pm$  0.195  &  0.053 &  3.0 &  1.8 & -44.5 &  2.4 &  0.7 & 131.4 &   \\
09:58:46.374$\pm$   0.010 & 2:16:02.36$\pm$      0.00  &  1.380$\pm$  0.156  &   4.522$\pm$  0.530  &  0.051 &  4.0 &  2.5 & -86.1 &  3.6 &  1.7 &  91.6 &   \\
09:58:48.461$\pm$ $\dots$ & 1:58:33.59$\pm$   $\dots$  &  0.350$\pm$ $\dots$ &   0.523$\pm$ $\dots$ &  0.070 &  0.0 &  0.0 &   0.0 &  0.0 &  0.0 &   0.0 & y \\
09:58:51.596$\pm$   0.006 & 2:08:58.11$\pm$      0.01  &  0.252$\pm$  0.071  &   0.271$\pm$  0.104  &  0.047 &  2.3 &  1.4 & -45.2 &  1.4 &  0.0 & 126.9 &   \\
09:58:55.618$\pm$   0.003 & 2:00:56.19$\pm$      0.00  &  0.257$\pm$  0.075  &   0.297$\pm$  0.115  &  0.049 &  2.3 &  1.6 &  26.6 &  1.5 &  0.0 &  36.4 &   \\
09:58:58.285$\pm$   0.003 & 2:10:46.37$\pm$      0.00  &  0.675$\pm$  0.092  &   0.870$\pm$  0.139  &  0.043 &  2.1 &  1.9 & -44.9 &  1.1 &  0.7 &  91.8 &   \\
09:58:58.353$\pm$   0.005 & 2:12:49.76$\pm$      0.00  &  0.230$\pm$  0.061  &   0.169$\pm$  0.065  &  0.042 &  1.8 &  1.2 & -64.6 &  0.5 &  0.0 &  99.5 &   \\
09:58:58.539$\pm$   0.001 & 2:14:59.09$\pm$      0.00  &  1.034$\pm$  0.121  &   1.105$\pm$  0.145  &  0.044 &  1.9 &  1.7 & -57.7 &  0.8 &  0.0 &  87.1 &   \\
09:58:58.996$\pm$   0.005 & 1:56:46.03$\pm$      0.01  &  0.295$\pm$  0.084  &   0.294$\pm$  0.115  &  0.056 &  2.0 &  1.5 & -18.5 &  0.6 &  0.0 & 168.0 &   \\
09:58:59.344$\pm$   0.004 & 2:20:44.45$\pm$      0.00  &  0.366$\pm$  0.072  &   0.397$\pm$  0.101  &  0.043 &  2.1 &  1.6 & -65.5 &  1.2 &  0.0 &  99.1 &   \\
09:58:59.354$\pm$   0.005 & 2:01:27.92$\pm$      0.01  &  0.222$\pm$  0.064  &   0.205$\pm$  0.082  &  0.043 &  1.7 &  1.7 & -51.6 &  0.6 &  0.0 & 157.0 &   \\
09:59:01.843$\pm$   0.003 & 2:14:49.90$\pm$      0.00  &  0.228$\pm$  0.062  &   0.205$\pm$  0.077  &  0.041 &  2.0 &  1.4 & -19.0 &  0.6 &  0.0 & 165.2 &   \\
09:59:06.274$\pm$   0.001 & 1:59:47.72$\pm$      0.00  &  0.224$\pm$  0.065  &   0.191$\pm$  0.078  &  0.044 &  1.7 &  1.5 &   4.3 &  0.0 &  0.0 &   0.0 &   \\
09:59:07.211$\pm$   0.005 & 2:16:30.14$\pm$      0.01  &  0.198$\pm$  0.056  &   0.191$\pm$  0.074  &  0.037 &  2.0 &  1.5 & -14.5 &  0.7 &  0.0 & 176.7 &   \\
09:59:07.475$\pm$   0.003 & 1:55:55.73$\pm$      0.00  &  0.589$\pm$  0.090  &   0.907$\pm$  0.163  &  0.046 &  2.3 &  2.1 &  37.9 &  1.6 &  1.0 &  53.9 &   \\
09:59:09.194$\pm$   0.005 & 1:47:42.96$\pm$      0.01  &  0.422$\pm$  0.120  &   0.323$\pm$  0.131  &  0.082 &  1.9 &  1.3 & -26.7 &  0.1 &  0.0 & 148.9 &   \\
09:59:10.305$\pm$   0.024 & 2:07:32.37$\pm$      0.01  &  0.220$\pm$  0.056  &   0.470$\pm$  0.143  &  0.035 &  3.2 &  2.0 & -63.2 &  2.7 &  1.0 & 112.0 &   \\
09:59:10.819$\pm$   0.004 & 2:19:09.93$\pm$      0.00  &  0.224$\pm$  0.057  &   0.284$\pm$  0.094  &  0.036 &  2.1 &  1.8 & -15.7 &  0.9 &  0.8 &  14.2 &   \\
09:59:11.161$\pm$   0.016 & 2:28:31.73$\pm$      0.01  &  0.400$\pm$  0.081  &   0.616$\pm$  0.155  &  0.048 &  2.4 &  1.9 & -61.9 &  1.6 &  0.7 & 104.1 &   \\
09:59:13.295$\pm$   0.003 & 1:52:28.25$\pm$      0.00  &  0.651$\pm$  0.110  &   0.862$\pm$  0.180  &  0.061 &  2.2 &  1.8 &  53.4 &  1.5 &  0.0 &  58.8 &   \\
09:59:13.882$\pm$   0.008 & 2:08:37.00$\pm$      0.00  &  0.222$\pm$  0.056  &   0.313$\pm$  0.100  &  0.035 &  2.6 &  1.7 & -80.2 &  2.0 &  0.0 &  93.5 &   \\
09:59:14.736$\pm$   0.003 & 2:12:19.54$\pm$      0.00  &  0.700$\pm$  0.084  &   0.859$\pm$  0.115  &  0.032 &  2.1 &  1.8 & -28.0 &  0.9 &  0.8 & 125.4 &   \\
09:59:14.795$\pm$   0.003 & 2:12:44.33$\pm$      0.00  &  0.214$\pm$  0.051  &   0.221$\pm$  0.072  &  0.033 &  1.9 &  1.6 & -27.8 &  0.3 &  0.0 & 109.6 &   \\
09:59:15.282$\pm$   0.043 & 2:01:23.99$\pm$      0.05  &  0.184$\pm$  0.056  &   0.422$\pm$  0.154  &  0.036 &  2.9 &  2.4 &  30.0 &  2.4 &  1.6 &  39.5 &   \\
09:59:15.464$\pm$   0.013 & 2:09:05.43$\pm$      0.01  &  0.176$\pm$  0.054  &   0.225$\pm$  0.090  &  0.035 &  2.3 &  1.7 & -78.0 &  1.6 &  0.0 &  92.2 &   \\
09:59:16.516$\pm$   0.003 & 2:09:44.35$\pm$      0.00  &  0.263$\pm$  0.055  &   0.284$\pm$  0.078  &  0.034 &  2.2 &  1.5 & -33.0 &  1.1 &  0.0 & 140.7 &   \\
09:59:17.751$\pm$   0.002 & 2:09:27.92$\pm$      0.00  &  0.367$\pm$  0.062  &   0.477$\pm$  0.099  &  0.034 &  2.1 &  1.9 &  -5.5 &  1.1 &  0.8 &  42.3 &   \\
09:59:18.713$\pm$   0.000 & 2:09:51.45$\pm$      0.00  &  1.228$\pm$  0.132  &   1.323$\pm$  0.151  &  0.034 &  2.0 &  1.6 & -28.2 &  0.6 &  0.0 & 139.3 &   \\
09:59:19.456$\pm$   0.002 & 2:34:12.98$\pm$      0.00  &  0.387$\pm$  0.105  &   0.886$\pm$  0.284  &  0.066 &  2.8 &  2.5 & -23.7 &  2.1 &  1.9 & 154.9 &   \\
09:59:29.230$\pm$ $\dots$ & 2:28:45.11$\pm$   $\dots$  &  0.198$\pm$ $\dots$ &   0.613$\pm$ $\dots$ &  0.040 &  0.0 &  0.0 &   0.0 &  0.0 &  0.0 &   0.0 & y \\
09:59:27.204$\pm$   0.009 & 2:37:38.04$\pm$      0.01  &  0.602$\pm$  0.119  &   3.002$\pm$  0.635  &  0.074 &  4.1 &  3.7 &   2.7 &  3.7 &  3.3 &  11.9 &   \\
09:59:24.028$\pm$   0.004 & 2:27:07.97$\pm$      0.00  &  0.750$\pm$  0.093  &   0.970$\pm$  0.137  &  0.038 &  2.2 &  1.8 & -36.6 &  1.2 &  0.7 & 125.5 &   \\
09:59:30.009$\pm$   0.011 & 2:03:45.45$\pm$      0.01  &  0.149$\pm$  0.048  &   0.212$\pm$  0.088  &  0.031 &  2.2 &  1.9 &  76.9 &  1.5 &  0.1 &  72.3 &   \\
09:59:30.903$\pm$   0.004 & 2:32:09.48$\pm$      0.00  &  0.254$\pm$  0.070  &   0.225$\pm$  0.087  &  0.047 &  1.9 &  1.4 & -45.1 &  0.5 &  0.0 & 115.3 &   \\
09:59:32.504$\pm$   0.003 & 2:10:37.75$\pm$      0.00  &  0.315$\pm$  0.052  &   0.443$\pm$  0.088  &  0.028 &  2.3 &  1.9 & -21.9 &  1.3 &  1.0 & 159.9 &   \\
09:59:33.387$\pm$   0.002 & 2:14:22.51$\pm$      0.00  &  0.140$\pm$  0.040  &   0.129$\pm$  0.052  &  0.027 &  1.9 &  1.4 &  24.9 &  0.8 &  0.0 &  40.3 &   \\
09:59:34.709$\pm$   0.004 & 2:12:29.05$\pm$      0.01  &  0.265$\pm$  0.049  &   0.535$\pm$  0.117  &  0.028 &  2.9 &  2.1 &  -1.3 &  2.2 &  1.3 &   5.0 &   \\
09:59:34.760$\pm$   0.004 & 2:06:33.67$\pm$      0.01  &  0.172$\pm$  0.042  &   0.152$\pm$  0.052  &  0.028 &  2.0 &  1.3 & -31.6 &  0.7 &  0.0 & 141.7 &   \\
09:59:34.815$\pm$   0.001 & 2:37:45.26$\pm$      0.00  &  1.803$\pm$  0.205  &   2.539$\pm$  0.312  &  0.067 &  2.7 &  1.6 & -22.9 &  1.9 &  0.0 & 157.1 &   \\
09:59:35.736$\pm$   0.005 & 1:58:05.37$\pm$      0.00  &  0.263$\pm$  0.055  &   0.341$\pm$  0.090  &  0.033 &  2.1 &  1.9 & -57.6 &  1.2 &  0.6 &  89.2 &   \\
09:59:36.222$\pm$   0.012 & 2:08:08.01$\pm$      0.01  &  0.209$\pm$  0.046  &   0.329$\pm$  0.090  &  0.028 &  2.5 &  1.9 & -53.3 &  1.7 &  0.8 & 115.1 &   \\
09:59:37.419$\pm$   0.000 & 2:23:47.14$\pm$      0.00  &  0.955$\pm$  0.107  &   1.189$\pm$  0.145  &  0.034 &  2.2 &  1.7 & -26.7 &  1.1 &  0.6 & 149.1 &   \\
09:59:40.748$\pm$   0.003 & 2:19:38.75$\pm$      0.00  &  0.304$\pm$  0.051  &   0.348$\pm$  0.074  &  0.029 &  2.0 &  1.7 &  15.7 &  1.0 &  0.0 &  40.4 &   \\
09:59:43.922$\pm$   0.002 & 2:33:32.37$\pm$      0.00  &  0.312$\pm$  0.074  &   0.318$\pm$  0.103  &  0.048 &  2.2 &  1.4 &  -7.5 &  1.2 &  0.0 &   0.1 &   \\
09:59:45.191$\pm$   0.002 & 2:34:39.22$\pm$      0.00  &  0.306$\pm$  0.078  &   0.322$\pm$  0.112  &  0.051 &  2.1 &  1.5 & -23.4 &  0.9 &  0.0 & 156.2 &   \\
09:59:45.686$\pm$   0.007 & 2:16:09.93$\pm$      0.01  &  0.154$\pm$  0.041  &   0.187$\pm$  0.065  &  0.026 &  2.1 &  1.8 &  15.3 &  1.2 &  0.2 &  39.2 &   \\
09:59:46.194$\pm$ $\dots$ & 2:36:03.19$\pm$   $\dots$  &  1.925$\pm$ $\dots$ &  14.367$\pm$ $\dots$ &  0.056 &  0.0 &  0.0 &   0.0 &  0.0 &  0.0 &   0.0 & y \\
09:59:47.869$\pm$   0.002 & 2:10:25.39$\pm$      0.00  &  0.163$\pm$  0.040  &   0.205$\pm$  0.065  &  0.025 &  2.2 &  1.7 &  14.5 &  1.3 &  0.0 &  30.1 &   \\
09:59:48.685$\pm$   0.002 & 2:23:10.74$\pm$      0.00  &  0.703$\pm$  0.083  &   0.817$\pm$  0.107  &  0.030 &  2.1 &  1.7 & -24.3 &  0.9 &  0.6 & 152.8 &   \\
09:59:48.838$\pm$   0.003 & 2:12:44.78$\pm$      0.00  &  0.397$\pm$  0.054  &   0.495$\pm$  0.079  &  0.025 &  2.3 &  1.7 & -33.1 &  1.3 &  0.5 & 139.7 &   \\
09:59:49.597$\pm$   0.001 & 1:59:44.17$\pm$      0.00  &  0.207$\pm$  0.047  &   0.284$\pm$  0.082  &  0.029 &  2.2 &  1.9 & -38.5 &  1.2 &  0.9 & 112.2 &   \\
09:59:49.912$\pm$ $\dots$ & 1:56:50.37$\pm$   $\dots$  &  0.605$\pm$ $\dots$ &   1.492$\pm$ $\dots$ &  0.032 &  0.0 &  0.0 &   0.0 &  0.0 &  0.0 &   0.0 & y \\
09:59:50.265$\pm$   0.001 & 1:48:05.12$\pm$      0.00  &  0.732$\pm$  0.103  &   1.026$\pm$  0.170  &  0.050 &  2.5 &  1.7 &   3.4 &  1.7 &  0.2 &  11.9 &   \\
09:59:50.908$\pm$   0.006 & 2:09:34.36$\pm$      0.01  &  0.132$\pm$  0.038  &   0.149$\pm$  0.058  &  0.025 &  2.0 &  1.8 &  34.3 &  1.1 &  0.0 &  53.6 &   \\
09:59:51.932$\pm$   0.000 & 2:05:42.55$\pm$      0.00  &  1.012$\pm$  0.108  &   1.112$\pm$  0.125  &  0.026 &  2.0 &  1.7 & -23.1 &  0.6 &  0.6 & 155.2 &   \\
09:59:52.220$\pm$   0.005 & 2:09:56.11$\pm$      0.01  &  0.126$\pm$  0.039  &   0.208$\pm$  0.080  &  0.025 &  2.7 &  1.9 & -17.5 &  1.9 &  1.0 & 164.7 &   \\
09:59:52.501$\pm$   0.015 & 2:07:51.03$\pm$      0.01  &  0.136$\pm$  0.041  &   0.243$\pm$  0.090  &  0.026 &  2.9 &  1.9 & -41.1 &  2.2 &  1.0 & 134.4 &   \\
09:59:54.247$\pm$   0.001 & 2:24:38.01$\pm$      0.00  &  0.379$\pm$  0.057  &   0.450$\pm$  0.083  &  0.030 &  2.0 &  1.8 & -74.1 &  1.1 &  0.0 &  82.8 &   \\
09:59:54.691$\pm$   0.000 & 2:30:15.31$\pm$      0.00  &  1.290$\pm$  0.140  &   1.380$\pm$  0.160  &  0.038 &  1.9 &  1.7 & -35.2 &  0.6 &  0.0 & 175.5 &   \\
09:59:55.975$\pm$   0.009 & 2:19:01.35$\pm$      0.01  &  0.136$\pm$  0.040  &   0.188$\pm$  0.072  &  0.026 &  2.2 &  1.9 &  -1.7 &  1.3 &  0.8 &  26.9 &   \\
09:59:57.994$\pm$ $\dots$ & 2:18:09.72$\pm$   $\dots$  &  1.923$\pm$ $\dots$ &   6.596$\pm$ $\dots$ &  0.028 &  0.0 &  0.0 &   0.0 &  0.0 &  0.0 &   0.0 & y \\
09:59:58.527$\pm$   0.002 & 1:52:54.85$\pm$      0.00  &  0.262$\pm$  0.056  &   0.312$\pm$  0.086  &  0.034 &  2.0 &  1.8 & -13.7 &  0.9 &  0.6 &  48.8 &   \\
09:59:58.783$\pm$   0.023 & 2:07:14.99$\pm$      0.01  &  0.284$\pm$  0.047  &   1.061$\pm$  0.189  &  0.026 &  4.1 &  2.8 & -76.4 &  3.7 &  2.1 & 100.5 &   \\
09:59:58.941$\pm$   0.007 & 2:18:03.67$\pm$      0.01  &  0.141$\pm$  0.042  &   0.142$\pm$  0.058  &  0.028 &  1.9 &  1.6 & -63.8 &  0.8 &  0.0 &  91.6 &   \\
09:59:59.079$\pm$   0.001 & 1:48:36.61$\pm$      0.00  &  2.089$\pm$  0.222  &   4.557$\pm$  0.496  &  0.050 &  2.8 &  2.4 &  83.5 &  2.3 &  1.5 &  78.0 &   \\
09:59:59.247$\pm$   0.001 & 2:09:16.44$\pm$      0.00  &  0.171$\pm$  0.041  &   0.214$\pm$  0.067  &  0.026 &  2.1 &  1.8 & -25.1 &  0.9 &  0.8 & 138.9 &   \\
09:59:59.327$\pm$   0.005 & 1:48:40.23$\pm$      0.01  &  1.068$\pm$  0.130  &   2.522$\pm$  0.328  &  0.050 &  3.0 &  2.4 &  42.4 &  2.5 &  1.5 &  48.1 &   \\
10:00:00.616$\pm$   0.000 & 2:15:31.07$\pm$      0.00  &  3.717$\pm$  0.373  &   4.156$\pm$  0.419  &  0.025 &  2.0 &  1.7 & -22.0 &  0.6 &  0.6 & 173.6 &   \\
10:00:01.712$\pm$   0.003 & 2:27:12.58$\pm$      0.00  &  0.408$\pm$  0.062  &   0.483$\pm$  0.089  &  0.032 &  2.0 &  1.8 & -31.9 &  0.9 &  0.6 &  84.7 &   \\
10:00:02.217$\pm$   0.004 & 2:16:22.01$\pm$      0.00  &  0.338$\pm$  0.052  &   0.464$\pm$  0.086  &  0.027 &  2.1 &  2.0 &  61.3 &  1.4 &  0.6 &  65.4 &   \\
10:00:02.803$\pm$   0.001 & 1:46:36.04$\pm$      0.00  &  0.326$\pm$  0.086  &   0.353$\pm$  0.126  &  0.056 &  2.0 &  1.6 &   3.3 &  0.9 &  0.0 &  26.4 &   \\
10:00:02.866$\pm$   0.008 & 2:29:11.80$\pm$      0.01  &  0.228$\pm$  0.059  &   0.337$\pm$  0.110  &  0.037 &  2.3 &  2.0 &  -2.3 &  1.4 &  1.1 &  24.8 &   \\
10:00:03.400$\pm$   0.004 & 2:07:23.18$\pm$      0.01  &  0.628$\pm$  0.074  &   1.380$\pm$  0.172  &  0.026 &  3.0 &  2.2 & -27.7 &  2.3 &  1.5 & 150.7 &   \\
10:00:04.080$\pm$   0.003 & 2:34:25.65$\pm$      0.00  &  0.217$\pm$  0.067  &   0.233$\pm$  0.097  &  0.044 &  2.1 &  1.6 & -57.8 &  1.1 &  0.0 & 105.4 &   \\
10:00:05.359$\pm$   0.004 & 2:30:59.27$\pm$      0.00  &  0.251$\pm$  0.062  &   0.262$\pm$  0.087  &  0.040 &  2.2 &  1.5 & -46.9 &  1.2 &  0.0 & 121.9 &   \\
10:00:05.434$\pm$   0.000 & 2:30:29.06$\pm$      0.00  &  2.699$\pm$  0.276  &   3.671$\pm$  0.382  &  0.040 &  2.2 &  1.9 & -42.0 &  1.2 &  0.9 & 109.0 &   \\
10:00:06.175$\pm$   0.001 & 2:40:00.34$\pm$      0.00  &  1.330$\pm$  0.171  &   2.164$\pm$  0.311  &  0.073 &  2.8 &  1.8 &  -6.7 &  2.1 &  0.8 & 177.6 &   \\
10:00:06.961$\pm$   0.001 & 2:17:33.80$\pm$      0.00  &  0.275$\pm$  0.047  &   0.366$\pm$  0.077  &  0.026 &  2.2 &  1.8 &  -0.4 &  1.2 &  0.6 &  20.0 &   \\
10:00:07.226$\pm$   0.002 & 2:40:49.02$\pm$      0.00  &  1.356$\pm$  0.184  &   3.135$\pm$  0.468  &  0.084 &  3.0 &  2.3 &   1.6 &  2.4 &  1.6 &   9.0 &   \\
10:00:07.430$\pm$   0.010 & 2:40:51.24$\pm$      0.01  &  0.682$\pm$  0.141  &   1.825$\pm$  0.432  &  0.084 &  3.1 &  2.6 & -10.1 &  2.5 &  2.0 & 176.6 &   \\
10:00:08.921$\pm$   0.003 & 2:14:40.56$\pm$      0.00  &  0.175$\pm$  0.040  &   0.190$\pm$  0.057  &  0.025 &  2.0 &  1.6 &  -1.3 &  0.8 &  0.0 &  22.1 &   \\
10:00:09.495$\pm$   0.005 & 2:22:19.48$\pm$      0.01  &  0.203$\pm$  0.044  &   0.258$\pm$  0.072  &  0.027 &  2.0 &  1.9 & -12.5 &  1.0 &  0.6 &  61.2 &   \\
10:00:10.156$\pm$   0.005 & 2:41:41.40$\pm$      0.02  &  0.906$\pm$  0.164  &   1.899$\pm$  0.403  &  0.094 &  3.3 &  1.9 &  -0.5 &  2.7 &  0.9 &   2.7 &   \\
10:00:10.345$\pm$   0.006 & 1:50:38.54$\pm$      0.01  &  0.335$\pm$  0.066  &   0.408$\pm$  0.102  &  0.039 &  2.0 &  1.9 &  76.6 &  1.2 &  0.1 &  69.6 &   \\
10:00:10.992$\pm$   0.023 & 2:07:23.76$\pm$      0.04  &  0.128$\pm$  0.039  &   0.245$\pm$  0.092  &  0.025 &  3.0 &  1.9 & -15.9 &  2.3 &  1.0 & 165.8 &   \\
10:00:12.092$\pm$   0.002 & 2:08:20.27$\pm$      0.00  &  0.324$\pm$  0.049  &   0.438$\pm$  0.079  &  0.025 &  2.2 &  1.8 & -35.9 &  1.2 &  0.7 & 126.6 &   \\
10:00:13.487$\pm$   0.001 & 2:18:15.55$\pm$      0.00  &  0.160$\pm$  0.039  &   0.144$\pm$  0.049  &  0.026 &  2.1 &  1.3 & -59.0 &  1.1 &  0.0 & 109.7 &   \\
10:00:13.583$\pm$   0.005 & 2:12:30.69$\pm$      0.01  &  0.160$\pm$  0.042  &   0.166$\pm$  0.058  &  0.027 &  2.0 &  1.6 & -16.9 &  0.7 &  0.0 & 177.2 &   \\
10:00:13.744$\pm$   0.001 & 2:12:21.35$\pm$      0.00  &  0.265$\pm$  0.047  &   0.292$\pm$  0.066  &  0.027 &  2.0 &  1.7 &  -2.9 &  0.9 &  0.0 &  29.9 &   \\
10:00:14.191$\pm$   0.001 & 2:13:11.99$\pm$      0.00  &  0.738$\pm$  0.083  &   1.267$\pm$  0.152  &  0.026 &  2.5 &  2.0 & -22.6 &  1.6 &  1.2 & 157.8 &   \\
10:00:15.303$\pm$   0.007 & 2:12:40.62$\pm$      0.01  &  0.184$\pm$  0.044  &   0.332$\pm$  0.096  &  0.027 &  2.7 &  2.1 & -12.0 &  1.9 &  1.3 & 173.8 &   \\
10:00:16.064$\pm$   0.003 & 2:12:37.35$\pm$      0.00  &  0.199$\pm$  0.045  &   0.355$\pm$  0.097  &  0.027 &  2.6 &  2.1 &  -7.2 &  1.8 &  1.3 &   3.2 &   \\
10:00:16.575$\pm$   0.000 & 2:26:38.37$\pm$      0.00  &  4.177$\pm$  0.420  &   4.743$\pm$  0.479  &  0.031 &  2.0 &  1.7 & -28.6 &  0.7 &  0.5 & 117.1 &   \\
10:00:16.840$\pm$   0.001 & 1:51:32.86$\pm$      0.00  &  0.663$\pm$  0.085  &   0.915$\pm$  0.135  &  0.037 &  2.1 &  2.0 & -25.6 &  1.2 &  0.9 &  68.7 &   \\
10:00:17.167$\pm$   0.014 & 2:09:29.32$\pm$      0.02  &  0.152$\pm$  0.039  &   0.204$\pm$  0.068  &  0.025 &  2.5 &  1.6 & -32.9 &  1.6 &  0.0 & 143.4 &   \\
10:00:17.797$\pm$   0.006 & 1:51:21.03$\pm$      0.00  &  0.185$\pm$  0.055  &   0.181$\pm$  0.075  &  0.037 &  1.8 &  1.6 & -56.3 &  0.6 &  0.0 &  86.3 &   \\
10:00:18.510$\pm$   0.002 & 2:32:56.54$\pm$      0.00  &  0.410$\pm$  0.074  &   0.950$\pm$  0.198  &  0.042 &  3.1 &  2.3 &  21.6 &  2.6 &  1.5 &  28.5 &   \\
10:00:18.757$\pm$   0.006 & 1:53:54.64$\pm$      0.00  &  0.770$\pm$  0.092  &   1.071$\pm$  0.141  &  0.034 &  2.1 &  2.0 &  26.7 &  1.3 &  0.7 &  57.0 &   \\
10:00:19.207$\pm$   0.008 & 2:13:36.72$\pm$      0.01  &  0.152$\pm$  0.041  &   0.197$\pm$  0.069  &  0.026 &  2.2 &  1.8 & -22.0 &  1.1 &  0.8 & 159.9 &   \\
10:00:21.771$\pm$   0.001 & 2:12:20.12$\pm$      0.00  &  0.304$\pm$  0.047  &   0.315$\pm$  0.060  &  0.025 &  2.0 &  1.6 & -18.6 &  0.6 &  0.0 & 172.3 &   \\
10:00:21.778$\pm$   0.001 & 2:00:00.22$\pm$      0.00  &  0.675$\pm$  0.079  &   1.085$\pm$  0.138  &  0.028 &  2.6 &  1.9 &  -5.9 &  1.8 &  1.0 &   1.4 &   \\
10:00:21.794$\pm$   0.002 & 2:23:28.81$\pm$      0.00  &  0.193$\pm$  0.047  &   0.240$\pm$  0.077  &  0.030 &  2.0 &  1.9 & -50.5 &  1.1 &  0.5 &  77.6 &   \\
10:00:22.906$\pm$   0.008 & 2:23:04.64$\pm$      0.01  &  0.148$\pm$  0.046  &   0.202$\pm$  0.082  &  0.030 &  2.2 &  1.9 & -65.6 &  1.4 &  0.6 &  93.2 &   \\
10:00:22.971$\pm$   0.007 & 2:13:12.70$\pm$      0.01  &  0.128$\pm$  0.037  &   0.359$\pm$  0.121  &  0.024 &  3.5 &  2.4 & -30.1 &  2.9 &  1.8 & 148.6 &   \\
10:00:23.440$\pm$   0.001 & 2:06:39.09$\pm$      0.00  &  0.131$\pm$  0.038  &   0.146$\pm$  0.057  &  0.025 &  1.9 &  1.8 & -49.5 &  0.9 &  0.0 & 166.8 &   \\
10:00:23.836$\pm$   0.002 & 2:01:05.54$\pm$      0.01  &  0.188$\pm$  0.042  &   0.184$\pm$  0.056  &  0.027 &  2.2 &  1.4 &  -5.1 &  1.2 &  0.0 &   3.3 &   \\
10:00:24.075$\pm$   0.086 & 2:31:02.70$\pm$      0.06  &  0.154$\pm$  0.060  &   0.370$\pm$  0.171  &  0.039 &  2.9 &  2.5 & -80.4 &  2.4 &  1.7 &  89.5 &   \\
10:00:25.477$\pm$   0.001 & 2:00:51.80$\pm$      0.00  &  0.726$\pm$  0.082  &   0.896$\pm$  0.111  &  0.027 &  2.2 &  1.7 & -22.3 &  1.1 &  0.6 & 158.5 &   \\
10:00:25.607$\pm$   0.001 & 2:03:16.17$\pm$      0.00  &  0.308$\pm$  0.049  &   0.380$\pm$  0.074  &  0.026 &  2.2 &  1.7 & -23.0 &  1.1 &  0.6 & 157.0 &   \\
10:00:25.800$\pm$   0.015 & 1:43:59.37$\pm$      0.03  &  0.361$\pm$  0.110  &   0.487$\pm$  0.192  &  0.071 &  2.3 &  1.8 &  -8.6 &  1.3 &  0.7 &   3.5 &   \\
10:00:27.659$\pm$ $\dots$ & 2:21:21.03$\pm$   $\dots$  &  0.251$\pm$ $\dots$ &   6.672$\pm$ $\dots$ &  0.034 &  0.0 &  0.0 &   0.0 &  0.0 &  0.0 &   0.0 & y \\
10:00:27.642$\pm$   0.003 & 2:01:03.19$\pm$      0.00  &  0.141$\pm$  0.042  &   0.185$\pm$  0.072  &  0.027 &  2.2 &  1.8 & -19.7 &  1.1 &  0.8 & 166.4 &   \\
10:00:28.549$\pm$   0.002 & 2:27:25.86$\pm$      0.00  &  0.418$\pm$  0.072  &   0.605$\pm$  0.127  &  0.040 &  2.2 &  2.0 & -11.6 &  1.2 &  1.1 &  42.1 &   \\
10:00:29.416$\pm$ $\dots$ & 2:40:38.66$\pm$   $\dots$  &  1.976$\pm$ $\dots$ &  35.197$\pm$ $\dots$ &  0.100 &  0.0 &  0.0 &   0.0 &  0.0 &  0.0 &   0.0 & y \\
10:00:28.951$\pm$   0.003 & 2:30:56.14$\pm$      0.00  &  0.195$\pm$  0.056  &   0.160$\pm$  0.065  &  0.038 &  1.9 &  1.3 & -30.3 &  0.2 &  0.0 & 141.5 &   \\
10:00:31.820$\pm$   0.004 & 2:12:43.42$\pm$      0.00  &  0.125$\pm$  0.036  &   0.122$\pm$  0.048  &  0.024 &  1.9 &  1.6 & -45.5 &  0.6 &  0.0 & 100.8 &   \\
10:00:34.057$\pm$   0.015 & 2:38:23.86$\pm$      0.02  &  0.287$\pm$  0.091  &   0.564$\pm$  0.220  &  0.059 &  3.1 &  1.9 & -36.5 &  2.5 &  1.0 & 140.8 &   \\
10:00:34.374$\pm$   0.004 & 2:21:21.78$\pm$      0.00  &  0.366$\pm$  0.057  &   0.451$\pm$  0.086  &  0.030 &  2.2 &  1.7 & -39.4 &  1.2 &  0.4 & 126.6 &   \\
10:00:35.700$\pm$   0.009 & 2:29:13.05$\pm$      0.02  &  0.175$\pm$  0.055  &   0.242$\pm$  0.100  &  0.036 &  2.3 &  1.8 &  19.3 &  1.5 &  0.3 &  33.4 &   \\
10:00:36.020$\pm$   0.010 & 2:39:37.77$\pm$      0.02  &  0.441$\pm$  0.106  &   0.826$\pm$  0.242  &  0.066 &  3.1 &  1.8 & -32.5 &  2.5 &  0.8 & 145.7 &   \\
10:00:36.051$\pm$   0.005 & 2:28:30.60$\pm$      0.01  &  0.316$\pm$  0.059  &   0.598$\pm$  0.133  &  0.034 &  2.7 &  2.2 & -39.0 &  2.0 &  1.5 & 131.2 &   \\
10:00:38.022$\pm$   0.008 & 2:08:22.85$\pm$      0.02  &  0.171$\pm$  0.042  &   0.272$\pm$  0.083  &  0.026 &  2.5 &  1.9 &   1.9 &  1.7 &  0.9 &  13.0 &   \\
10:00:41.419$\pm$   0.003 & 2:31:24.11$\pm$      0.00  &  0.465$\pm$  0.074  &   0.605$\pm$  0.118  &  0.040 &  2.1 &  1.9 & -28.5 &  1.0 &  0.9 &  82.0 &   \\
10:00:42.318$\pm$   0.011 & 2:00:50.30$\pm$      0.02  &  0.154$\pm$  0.043  &   0.346$\pm$  0.114  &  0.027 &  3.1 &  2.2 & -32.0 &  2.5 &  1.5 & 145.5 &   \\
10:00:43.177$\pm$ $\dots$ & 1:46:07.89$\pm$   $\dots$  &  4.633$\pm$ $\dots$ &  65.099$\pm$ $\dots$ &  0.070 &  0.0 &  0.0 &   0.0 &  0.0 &  0.0 &   0.0 & y \\
10:00:43.528$\pm$   0.004 & 2:25:24.36$\pm$      0.00  &  0.405$\pm$  0.062  &   0.591$\pm$  0.108  &  0.032 &  2.3 &  1.9 & -21.4 &  1.3 &  1.0 & 161.3 &   \\
10:00:45.802$\pm$   0.009 & 2:01:18.94$\pm$      0.01  &  0.242$\pm$  0.048  &   0.471$\pm$  0.110  &  0.028 &  3.0 &  2.0 & -29.3 &  2.3 &  1.2 & 149.1 &   \\
10:00:46.916$\pm$   0.000 & 2:07:26.50$\pm$      0.00  &  1.498$\pm$  0.155  &   1.737$\pm$  0.184  &  0.027 &  2.0 &  1.8 & -21.1 &  0.8 &  0.6 &  62.1 &   \\
10:00:47.599$\pm$ $\dots$ & 1:59:00.58$\pm$   $\dots$  &  2.065$\pm$ $\dots$ &  13.946$\pm$ $\dots$ &  0.033 &  0.0 &  0.0 &   0.0 &  0.0 &  0.0 &   0.0 & y \\
10:00:47.533$\pm$   0.005 & 2:09:41.05$\pm$      0.01  &  0.196$\pm$  0.041  &   0.213$\pm$  0.058  &  0.025 &  2.0 &  1.6 &  39.6 &  1.1 &  0.0 &  51.0 &   \\
10:00:47.570$\pm$ $\dots$ & 2:09:58.58$\pm$   $\dots$  &  0.138$\pm$ $\dots$ &   0.417$\pm$ $\dots$ &  0.025 &  0.0 &  0.0 &   0.0 &  0.0 &  0.0 &   0.0 & y \\
10:00:49.604$\pm$ $\dots$ & 1:49:23.88$\pm$   $\dots$  &  0.711$\pm$ $\dots$ &   4.720$\pm$ $\dots$ &  0.046 &  0.0 &  0.0 &   0.0 &  0.0 &  0.0 &   0.0 & y \\
10:00:48.883$\pm$   0.005 & 2:31:27.25$\pm$      0.00  &  0.293$\pm$  0.065  &   0.315$\pm$  0.093  &  0.041 &  1.8 &  1.8 &  55.1 &  0.8 &  0.0 &  67.0 &   \\
10:00:49.787$\pm$   0.001 & 2:16:54.79$\pm$      0.00  &  0.686$\pm$  0.080  &   0.768$\pm$  0.100  &  0.029 &  2.0 &  1.7 & -27.0 &  0.7 &  0.5 & 120.8 &   \\
10:00:49.912$\pm$   0.004 & 2:05:00.04$\pm$      0.01  &  0.167$\pm$  0.043  &   0.234$\pm$  0.077  &  0.027 &  2.3 &  1.8 & -28.7 &  1.3 &  0.8 & 145.6 &   \\
10:00:49.940$\pm$   0.006 & 2:22:44.91$\pm$      0.01  &  0.212$\pm$  0.047  &   0.340$\pm$  0.094  &  0.029 &  2.8 &  1.7 & -43.3 &  2.1 &  0.4 & 132.0 &   \\
10:00:50.043$\pm$ $\dots$ & 1:49:45.97$\pm$   $\dots$  &  0.390$\pm$ $\dots$ &   2.601$\pm$ $\dots$ &  0.043 &  0.0 &  0.0 &   0.0 &  0.0 &  0.0 &   0.0 & y \\
10:00:50.445$\pm$   0.002 & 2:33:56.26$\pm$      0.00  &  0.494$\pm$  0.084  &   0.762$\pm$  0.155  &  0.046 &  2.4 &  2.0 & -74.7 &  1.7 &  0.8 &  91.8 &   \\
10:00:50.571$\pm$   0.003 & 2:07:02.02$\pm$      0.01  &  0.232$\pm$  0.043  &   0.291$\pm$  0.068  &  0.025 &  2.4 &  1.6 & -26.8 &  1.5 &  0.0 & 151.4 &   \\
10:00:53.769$\pm$   0.012 & 2:16:14.85$\pm$      0.01  &  0.177$\pm$  0.045  &   0.273$\pm$  0.087  &  0.028 &  2.6 &  1.8 & -66.5 &  1.9 &  0.3 & 105.1 &   \\
10:00:55.107$\pm$   0.028 & 1:42:23.26$\pm$      0.02  &  0.455$\pm$  0.149  &   1.225$\pm$  0.470  &  0.097 &  3.0 &  2.7 & -67.5 &  2.5 &  2.0 &  96.6 &   \\
10:00:55.643$\pm$   0.001 & 1:56:45.52$\pm$      0.00  &  0.163$\pm$  0.045  &   0.137$\pm$  0.053  &  0.030 &  1.9 &  1.4 & -16.4 &  0.2 &  0.0 & 174.1 &   \\
10:00:56.086$\pm$   0.003 & 1:43:47.13$\pm$      0.01  &  0.412$\pm$  0.122  &   0.462$\pm$  0.184  &  0.080 &  2.1 &  1.6 &   5.8 &  1.1 &  0.0 &  23.1 &   \\
10:00:56.242$\pm$   0.003 & 2:01:21.32$\pm$      0.00  &  0.246$\pm$  0.046  &   0.299$\pm$  0.071  &  0.027 &  2.2 &  1.7 & -12.0 &  1.1 &  0.5 & 179.0 &   \\
10:00:56.635$\pm$   0.019 & 2:26:35.63$\pm$      0.03  &  0.156$\pm$  0.053  &   0.351$\pm$  0.142  &  0.034 &  3.0 &  2.3 & -12.1 &  2.3 &  1.6 & 172.0 &   \\
10:00:57.062$\pm$   0.036 & 2:29:42.53$\pm$      0.03  &  0.200$\pm$  0.063  &   0.278$\pm$  0.114  &  0.041 &  2.1 &  2.0 &  68.9 &  1.4 &  0.6 &  67.5 &   \\
10:00:57.098$\pm$   0.003 & 2:34:51.97$\pm$      0.01  &  0.391$\pm$  0.089  &   0.526$\pm$  0.153  &  0.055 &  2.3 &  1.8 & -19.1 &  1.3 &  0.8 & 164.9 &   \\
10:00:57.233$\pm$   0.021 & 2:03:21.54$\pm$      0.03  &  0.133$\pm$  0.040  &   0.182$\pm$  0.070  &  0.026 &  2.9 &  1.5 & -34.3 &  2.2 &  0.0 & 143.5 &   \\
10:00:57.942$\pm$   0.006 & 1:58:19.15$\pm$      0.01  &  0.211$\pm$  0.046  &   0.261$\pm$  0.073  &  0.028 &  2.4 &  1.6 &  -7.8 &  1.5 &  0.0 & 178.7 &   \\
10:00:58.069$\pm$ $\dots$ & 1:51:28.89$\pm$   $\dots$  &  2.135$\pm$ $\dots$ &  10.969$\pm$ $\dots$ &  0.045 &  0.0 &  0.0 &   0.0 &  0.0 &  0.0 &   0.0 & y \\
10:00:59.726$\pm$   0.022 & 2:06:48.26$\pm$      0.05  &  0.139$\pm$  0.042  &   0.402$\pm$  0.139  &  0.027 &  3.5 &  2.5 &  -2.1 &  3.0 &  1.9 &   1.7 &   \\
10:01:00.219$\pm$   0.004 & 1:51:50.38$\pm$      0.01  &  0.436$\pm$  0.082  &   0.638$\pm$  0.147  &  0.047 &  2.3 &  2.0 & -10.1 &  1.4 &  1.1 &  16.4 &   \\
10:01:00.332$\pm$   0.013 & 1:49:02.30$\pm$      0.02  &  0.262$\pm$  0.077  &   0.280$\pm$  0.112  &  0.051 &  2.2 &  1.5 & -28.8 &  1.1 &  0.0 & 147.3 &   \\
10:01:00.672$\pm$   0.005 & 2:16:41.12$\pm$      0.01  &  0.302$\pm$  0.051  &   0.560$\pm$  0.112  &  0.028 &  2.5 &  2.2 &  -0.5 &  1.7 &  1.4 &  23.5 &   \\
10:01:01.272$\pm$   0.001 & 2:01:18.01$\pm$      0.00  &  0.912$\pm$  0.100  &   1.188$\pm$  0.140  &  0.029 &  2.2 &  1.8 &  -9.9 &  1.2 &  0.7 &   7.7 &   \\
10:01:01.315$\pm$   0.029 & 2:21:59.03$\pm$      0.01  &  0.185$\pm$  0.048  &   0.412$\pm$  0.127  &  0.030 &  3.2 &  2.1 &  73.4 &  2.8 &  0.9 &  72.4 &   \\
10:01:02.384$\pm$   0.010 & 2:05:28.57$\pm$      0.02  &  0.515$\pm$  0.067  &   1.541$\pm$  0.213  &  0.029 &  3.5 &  2.6 &   6.2 &  3.0 &  2.0 &  11.3 &   \\
10:01:03.790$\pm$   0.030 & 1:55:33.48$\pm$      0.01  &  0.177$\pm$  0.059  &   0.311$\pm$  0.129  &  0.038 &  2.8 &  1.9 & -85.5 &  2.3 &  0.4 &  89.5 &   \\
10:01:04.521$\pm$   0.001 & 2:02:03.58$\pm$      0.00  &  1.162$\pm$  0.123  &   1.386$\pm$  0.153  &  0.028 &  2.0 &  1.8 & -34.0 &  0.9 &  0.6 &  86.8 &   \\
10:01:07.065$\pm$   0.016 & 2:22:39.05$\pm$      0.00  &  0.145$\pm$  0.046  &   0.197$\pm$  0.083  &  0.031 &  2.9 &  1.4 & -78.8 &  2.4 &  0.0 &  97.1 &   \\
10:01:07.169$\pm$   0.003 & 2:38:05.12$\pm$      0.00  &  0.429$\pm$  0.103  &   0.262$\pm$  0.091  &  0.071 &  1.5 &  1.3 & -42.7 &  0.0 &  0.0 &   0.0 &   \\
10:01:07.976$\pm$   0.003 & 2:22:25.96$\pm$      0.00  &  0.169$\pm$  0.047  &   0.133$\pm$  0.053  &  0.032 &  1.6 &  1.5 &  63.0 &  0.0 &  0.0 &   0.0 &   \\
10:01:08.992$\pm$   0.003 & 2:28:15.88$\pm$      0.00  &  0.444$\pm$  0.077  &   0.628$\pm$  0.133  &  0.043 &  2.4 &  1.8 &  -1.4 &  1.5 &  0.7 &   9.7 &   \\
10:01:09.243$\pm$   0.001 & 2:22:55.07$\pm$      0.00  &  0.476$\pm$  0.065  &   0.600$\pm$  0.097  &  0.031 &  2.4 &  1.6 & -19.4 &  1.5 &  0.0 & 162.3 &   \\
10:01:09.272$\pm$ $\dots$ & 2:17:20.83$\pm$   $\dots$  &  1.888$\pm$ $\dots$ &   3.688$\pm$ $\dots$ &  0.030 &  0.0 &  0.0 &   0.0 &  0.0 &  0.0 &   0.0 & y \\
10:01:10.630$\pm$   0.006 & 2:24:58.37$\pm$      0.01  &  0.406$\pm$  0.067  &   0.636$\pm$  0.125  &  0.036 &  2.2 &  2.2 &  24.9 &  1.5 &  1.1 &  67.0 &   \\
10:01:10.764$\pm$   0.002 & 2:02:04.05$\pm$      0.00  &  0.373$\pm$  0.059  &   0.453$\pm$  0.089  &  0.032 &  2.2 &  1.7 & -23.8 &  1.1 &  0.6 & 155.3 &   \\
10:01:11.710$\pm$   0.001 & 2:12:50.24$\pm$      0.00  &  0.143$\pm$  0.043  &   0.155$\pm$  0.062  &  0.028 &  2.1 &  1.6 & -25.9 &  0.9 &  0.0 & 150.3 &   \\
10:01:12.060$\pm$   0.002 & 2:41:06.75$\pm$      0.00  &  1.629$\pm$  0.214  &   3.194$\pm$  0.465  &  0.094 &  3.0 &  2.0 &   8.9 &  2.4 &  1.0 &  14.8 &   \\
10:01:12.967$\pm$   0.025 & 2:24:23.09$\pm$      0.04  &  0.148$\pm$  0.055  &   0.537$\pm$  0.225  &  0.037 &  3.7 &  3.0 &  -7.8 &  3.2 &  2.5 & 176.2 &   \\
10:01:13.593$\pm$   0.002 & 2:06:53.63$\pm$      0.00  &  0.398$\pm$  0.059  &   0.581$\pm$  0.102  &  0.030 &  2.2 &  2.1 &  -2.5 &  1.4 &  1.1 &  56.4 &   \\
10:01:13.526$\pm$ $\dots$ & 1:54:32.64$\pm$   $\dots$  &  1.034$\pm$ $\dots$ &   5.779$\pm$ $\dots$ &  0.045 &  0.0 &  0.0 &   0.0 &  0.0 &  0.0 &   0.0 & y \\
10:01:14.846$\pm$ $\dots$ & 2:02:08.54$\pm$   $\dots$  &  1.237$\pm$ $\dots$ &   4.046$\pm$ $\dots$ &  0.032 &  0.0 &  0.0 &   0.0 &  0.0 &  0.0 &   0.0 & y \\
10:01:15.510$\pm$   0.024 & 2:17:19.67$\pm$      0.02  &  0.188$\pm$  0.049  &   0.602$\pm$  0.177  &  0.031 &  3.4 &  2.8 &  49.5 &  3.0 &  2.1 &  53.2 &   \\
10:01:16.537$\pm$   0.001 & 1:50:50.42$\pm$      0.00  &  0.890$\pm$  0.113  &   0.890$\pm$  0.132  &  0.049 &  2.1 &  1.5 & -22.4 &  0.9 &  0.0 & 158.2 &   \\
10:01:17.207$\pm$   0.003 & 2:15:59.81$\pm$      0.00  &  0.159$\pm$  0.047  &   0.216$\pm$  0.082  &  0.030 &  2.2 &  1.9 &  19.3 &  1.4 &  0.6 &  40.7 &   \\
10:01:17.659$\pm$   0.001 & 1:44:16.12$\pm$      0.00  &  0.413$\pm$  0.122  &   0.402$\pm$  0.163  &  0.081 &  1.9 &  1.6 & -17.9 &  0.3 &  0.0 &  24.6 &   \\
10:01:17.967$\pm$   0.006 & 2:29:02.36$\pm$      0.01  &  0.335$\pm$  0.074  &   0.419$\pm$  0.120  &  0.046 &  2.4 &  1.6 & -33.8 &  1.5 &  0.0 & 141.3 &   \\
10:01:18.465$\pm$   0.006 & 2:05:38.77$\pm$      0.01  &  0.160$\pm$  0.045  &   0.197$\pm$  0.073  &  0.029 &  2.0 &  1.9 &  86.6 &  1.2 &  0.2 &  72.2 &   \\
10:01:19.572$\pm$   0.010 & 1:55:16.25$\pm$      0.01  &  0.387$\pm$  0.072  &   0.935$\pm$  0.198  &  0.041 &  3.0 &  2.5 & -76.8 &  2.5 &  1.7 &  94.2 &   \\
10:01:20.062$\pm$   0.000 & 2:34:43.76$\pm$      0.00  &  8.706$\pm$  0.876  &  10.738$\pm$  1.085  &  0.064 &  2.1 &  1.8 &  10.8 &  1.2 &  0.3 &  36.6 &   \\
10:01:22.455$\pm$   0.003 & 2:01:12.44$\pm$      0.00  &  2.776$\pm$  0.283  &   9.160$\pm$  0.938  &  0.037 &  4.0 &  2.5 &  42.5 &  3.6 &  1.7 &  44.7 &   \\
10:01:24.096$\pm$   0.001 & 2:17:06.37$\pm$      0.00  &  0.836$\pm$  0.095  &   1.657$\pm$  0.200  &  0.031 &  2.9 &  2.1 & -63.6 &  2.3 &  1.2 & 108.8 &   \\
10:01:24.037$\pm$   0.001 & 2:20:04.84$\pm$      0.00  &  0.500$\pm$  0.068  &   0.551$\pm$  0.089  &  0.032 &  2.1 &  1.6 & -31.2 &  0.9 &  0.0 & 139.1 &   \\
10:01:25.477$\pm$   0.001 & 2:28:40.08$\pm$      0.00  &  0.234$\pm$  0.066  &   0.208$\pm$  0.081  &  0.044 &  1.7 &  1.6 &  70.5 &  0.6 &  0.0 &  67.8 &   \\
10:01:27.991$\pm$   0.004 & 2:40:29.08$\pm$      0.01  &  0.816$\pm$  0.167  &   1.605$\pm$  0.394  &  0.099 &  2.9 &  2.0 &  32.5 &  2.4 &  0.8 &  38.3 &   \\
10:01:28.622$\pm$   0.005 & 1:57:30.58$\pm$      0.01  &  0.274$\pm$  0.059  &   0.291$\pm$  0.083  &  0.037 &  2.3 &  1.4 & -36.7 &  1.3 &  0.0 & 137.6 &   \\
10:01:28.868$\pm$   0.001 & 2:01:28.52$\pm$      0.00  &  0.274$\pm$  0.058  &   0.304$\pm$  0.085  &  0.036 &  2.0 &  1.7 & -46.2 &  0.9 &  0.0 & 102.3 &   \\
10:01:29.965$\pm$   0.013 & 2:17:05.15$\pm$      0.01  &  0.184$\pm$  0.050  &   0.221$\pm$  0.079  &  0.032 &  2.2 &  1.7 &  76.1 &  1.5 &  0.0 &  72.9 &   \\
10:01:30.744$\pm$   0.004 & 2:19:05.52$\pm$      0.00  &  0.542$\pm$  0.073  &   0.792$\pm$  0.122  &  0.033 &  2.4 &  1.8 & -61.9 &  1.6 &  0.4 & 106.1 &   \\
10:01:31.149$\pm$   0.000 & 2:29:24.74$\pm$      0.00  &  2.908$\pm$  0.300  &   4.156$\pm$  0.437  &  0.049 &  2.3 &  1.9 & -14.9 &  1.3 &  1.0 & 176.9 &   \\
10:01:31.229$\pm$   0.001 & 2:26:37.68$\pm$      0.00  &  3.433$\pm$  0.350  &   7.480$\pm$  0.769  &  0.045 &  2.8 &  2.4 &  15.6 &  2.2 &  1.6 &  30.1 &   \\
10:01:31.491$\pm$   0.001 & 2:26:40.53$\pm$      0.00  &  3.709$\pm$  0.377  &   8.432$\pm$  0.864  &  0.046 &  3.0 &  2.3 &  18.6 &  2.4 &  1.5 &  26.7 &   \\
10:01:32.137$\pm$   0.003 & 2:04:28.26$\pm$      0.01  &  0.331$\pm$  0.060  &   0.481$\pm$  0.107  &  0.034 &  2.4 &  1.8 & -17.1 &  1.5 &  0.8 & 167.0 &   \\
10:01:33.016$\pm$   0.012 & 2:21:09.78$\pm$      0.02  &  0.810$\pm$  0.098  &   0.882$\pm$  0.122  &  0.039 &  2.1 &  1.6 & -25.8 &  0.9 &  0.0 & 150.6 &   \\
10:01:33.474$\pm$   0.012 & 2:15:20.92$\pm$      0.02  &  0.215$\pm$  0.053  &   0.242$\pm$  0.080  &  0.034 &  2.0 &  1.7 & -40.9 &  0.8 &  0.2 & 105.5 &   \\
10:01:34.216$\pm$   0.001 & 2:09:17.51$\pm$      0.00  &  0.516$\pm$  0.071  &   0.522$\pm$  0.086  &  0.034 &  1.8 &  1.7 &  24.3 &  0.7 &  0.0 &  58.0 &   \\
10:01:34.332$\pm$   0.018 & 2:20:36.88$\pm$      0.06  &  0.226$\pm$  0.058  &   0.227$\pm$  0.079  &  0.038 &  2.2 &  1.4 & -27.3 &  1.1 &  0.0 & 150.3 &   \\
10:01:36.465$\pm$   0.018 & 2:26:41.57$\pm$      0.06  &  0.227$\pm$  0.067  &   0.882$\pm$  0.290  &  0.045 &  4.6 &  2.6 & -16.3 &  4.2 &  2.0 & 164.2 &   \\
10:01:36.699$\pm$   0.008 & 2:23:23.56$\pm$      0.00  &  0.231$\pm$  0.065  &   0.302$\pm$  0.111  &  0.042 &  2.3 &  1.8 & -60.2 &  1.5 &  0.4 & 104.9 &   \\
10:01:37.784$\pm$   0.001 & 1:48:11.73$\pm$      0.00  &  1.301$\pm$  0.171  &   2.150$\pm$  0.320  &  0.076 &  2.9 &  1.7 & -36.0 &  2.2 &  0.5 & 141.1 &   \\
10:01:38.373$\pm$   0.002 & 2:02:58.25$\pm$      0.00  &  0.237$\pm$  0.058  &   0.243$\pm$  0.079  &  0.037 &  2.0 &  1.6 & -12.8 &  0.7 &  0.0 &   6.6 &   \\
10:01:39.138$\pm$ $\dots$ & 1:51:41.03$\pm$   $\dots$  &  2.021$\pm$ $\dots$ &   7.092$\pm$ $\dots$ &  0.068 &  0.0 &  0.0 &   0.0 &  0.0 &  0.0 &   0.0 & y \\
10:01:39.747$\pm$   0.013 & 2:25:48.86$\pm$      0.01  &  0.255$\pm$  0.070  &   0.519$\pm$  0.172  &  0.044 &  2.8 &  2.2 &  70.1 &  2.3 &  1.1 &  69.3 &   \\
10:01:41.042$\pm$   0.018 & 1:59:03.86$\pm$      0.01  &  0.314$\pm$  0.069  &   0.928$\pm$  0.231  &  0.042 &  3.5 &  2.5 &  77.3 &  3.1 &  1.6 &  75.8 &   \\
10:01:41.442$\pm$   0.007 & 2:31:57.00$\pm$      0.01  &  0.457$\pm$  0.119  &   0.627$\pm$  0.209  &  0.075 &  2.4 &  1.8 &  53.1 &  1.8 &  0.0 &  57.1 &   \\
10:01:42.645$\pm$   0.007 & 2:07:52.87$\pm$      0.00  &  0.629$\pm$  0.081  &   0.766$\pm$  0.114  &  0.035 &  2.0 &  1.8 & -29.0 &  0.8 &  0.6 &  80.6 &   \\
10:01:43.455$\pm$   0.001 & 2:21:34.70$\pm$      0.00  &  1.102$\pm$  0.124  &   1.401$\pm$  0.170  &  0.039 &  2.2 &  1.8 & -31.2 &  1.1 &  0.8 & 135.5 &   \\
10:01:44.486$\pm$   0.023 & 2:13:46.09$\pm$      0.03  &  0.231$\pm$  0.060  &   0.414$\pm$  0.134  &  0.039 &  3.4 &  1.6 & -40.9 &  2.8 &  0.0 & 136.9 &   \\
10:01:44.824$\pm$   0.003 & 2:04:09.08$\pm$      0.00  &  0.652$\pm$  0.087  &   0.856$\pm$  0.134  &  0.040 &  2.2 &  1.8 & -37.2 &  1.2 &  0.7 & 124.6 &   \\
10:01:47.354$\pm$   0.000 & 2:03:14.13$\pm$      0.00  &  4.588$\pm$  0.463  &   5.998$\pm$  0.609  &  0.042 &  2.1 &  1.9 & -29.1 &  1.0 &  0.9 &  83.1 &   \\
10:01:47.307$\pm$   0.008 & 1:52:58.07$\pm$      0.00  &  0.508$\pm$  0.109  &   0.741$\pm$  0.200  &  0.066 &  2.4 &  1.8 & -78.4 &  1.7 &  0.0 &  92.2 &   \\
10:01:49.604$\pm$   0.005 & 2:33:34.97$\pm$      0.01  &  1.340$\pm$  0.184  &   2.964$\pm$  0.449  &  0.085 &  3.0 &  2.3 &  51.3 &  2.5 &  1.3 &  54.7 &   \\
10:01:51.620$\pm$   0.013 & 2:25:31.88$\pm$      0.02  &  0.293$\pm$  0.080  &   0.791$\pm$  0.252  &  0.051 &  3.0 &  2.7 & -40.1 &  2.4 &  2.1 & 122.4 &   \\
10:01:52.578$\pm$   0.009 & 2:19:54.33$\pm$      0.01  &  0.345$\pm$  0.069  &   0.506$\pm$  0.127  &  0.041 &  2.3 &  1.9 &  71.4 &  1.7 &  0.1 &  69.7 &   \\
10:01:53.456$\pm$   0.000 & 2:11:52.58$\pm$      0.00  &  2.231$\pm$  0.233  &   2.389$\pm$  0.258  &  0.046 &  2.0 &  1.6 & -22.5 &  0.6 &  0.0 & 158.8 &   \\
10:01:53.930$\pm$   0.009 & 2:05:38.60$\pm$      0.00  &  0.242$\pm$  0.065  &   0.324$\pm$  0.111  &  0.041 &  2.2 &  1.9 &  70.8 &  1.5 &  0.0 &  69.1 &   \\
10:01:54.092$\pm$   0.002 & 2:06:07.34$\pm$      0.00  &  0.327$\pm$  0.067  &   0.370$\pm$  0.099  &  0.041 &  2.2 &  1.6 & -29.9 &  1.1 &  0.0 & 144.5 &   \\
10:01:55.186$\pm$   0.003 & 2:27:42.19$\pm$      0.01  &  0.301$\pm$  0.089  &   0.308$\pm$  0.124  &  0.059 &  2.0 &  1.6 &  -8.9 &  0.7 &  0.0 &  13.1 &   \\
10:01:55.511$\pm$   0.002 & 2:03:58.44$\pm$      0.00  &  0.383$\pm$  0.075  &   0.556$\pm$  0.135  &  0.044 &  2.5 &  1.8 & -44.8 &  1.7 &  0.7 & 126.3 &   \\
10:02:00.741$\pm$   0.002 & 2:07:32.74$\pm$      0.00  &  0.243$\pm$  0.069  &   0.188$\pm$  0.077  &  0.048 &  1.6 &  1.4 &  20.7 &  0.0 &  0.0 &   0.0 &   \\
10:02:01.198$\pm$ $\dots$ & 2:13:24.27$\pm$   $\dots$  &  0.830$\pm$ $\dots$ &   4.697$\pm$ $\dots$ &  0.049 &  0.0 &  0.0 &   0.0 &  0.0 &  0.0 &   0.0 & y \\
10:02:02.564$\pm$   0.020 & 2:01:45.00$\pm$      0.03  &  0.285$\pm$  0.085  &   0.487$\pm$  0.179  &  0.054 &  2.7 &  2.0 & -27.7 &  1.9 &  1.2 & 150.1 &   \\
10:02:02.870$\pm$   0.014 & 2:00:26.27$\pm$      0.02  &  0.280$\pm$  0.088  &   0.727$\pm$  0.268  &  0.057 &  3.3 &  2.4 & -48.0 &  2.7 &  1.7 & 126.9 &   \\
10:02:05.467$\pm$   0.002 & 1:57:41.96$\pm$      0.00  &  0.420$\pm$  0.098  &   0.277$\pm$  0.093  &  0.067 &  1.6 &  1.2 &  -5.7 &  0.0 &  0.0 &   0.0 &   \\
10:02:08.542$\pm$   0.013 & 2:01:50.64$\pm$      0.01  &  0.332$\pm$  0.100  &   0.371$\pm$  0.150  &  0.066 &  2.3 &  1.5 &  60.6 &  1.6 &  0.0 &  62.2 &   \\
10:02:09.062$\pm$   0.001 & 2:16:02.50$\pm$      0.00  &  3.873$\pm$  0.395  &   4.977$\pm$  0.516  &  0.055 &  2.1 &  1.9 &  73.7 &  1.4 &  0.1 &  69.9 &   \\
10:02:09.152$\pm$   0.003 & 2:23:34.91$\pm$      0.00  &  0.381$\pm$  0.113  &   0.361$\pm$  0.149  &  0.076 &  1.9 &  1.5 & -57.3 &  0.7 &  0.0 & 100.2 &   \\
10:02:09.299$\pm$   0.005 & 2:00:55.56$\pm$      0.00  &  0.383$\pm$  0.119  &   0.452$\pm$  0.186  &  0.078 &  2.2 &  1.7 &  60.7 &  1.5 &  0.0 &  62.9 &   \\
10:02:10.109$\pm$   0.005 & 2:16:37.99$\pm$      0.00  &  0.477$\pm$  0.097  &   0.654$\pm$  0.168  &  0.058 &  2.2 &  1.9 & -58.7 &  1.3 &  0.7 &  97.3 &   \\
10:02:10.333$\pm$   0.003 & 2:03:49.77$\pm$      0.01  &  0.312$\pm$  0.091  &   0.282$\pm$  0.116  &  0.062 &  2.1 &  1.3 &  23.7 &  1.2 &  0.0 &  34.0 &   \\
10:02:20.898$\pm$   0.009 & 2:22:21.00$\pm$      0.04  &  0.400$\pm$  0.125  &   0.801$\pm$  0.307  &  0.082 &  3.4 &  1.8 & -13.9 &  2.8 &  0.8 & 167.4 &   \\
10:02:24.144$\pm$   0.001 & 2:16:21.35$\pm$      0.00  &  6.249$\pm$  0.633  &   9.730$\pm$  0.994  &  0.069 &  2.5 &  1.9 &  75.9 &  1.9 &  0.1 &  73.4 &   \\
10:02:27.036$\pm$   0.002 & 2:21:19.37$\pm$      0.00  &  0.575$\pm$  0.146  &   0.795$\pm$  0.259  &  0.093 &  2.7 &  1.6 &  81.1 &  2.2 &  0.0 &  78.6 &   \\
10:02:28.775$\pm$   0.002 & 2:17:21.91$\pm$      0.00  &  1.804$\pm$  0.216  &   3.418$\pm$  0.445  &  0.081 &  2.6 &  2.2 &  81.4 &  2.0 &  1.1 &  76.4 &   \\
10:02:29.728$\pm$ $\dots$ & 2:09:09.87$\pm$   $\dots$  &  0.691$\pm$ $\dots$ &   3.931$\pm$ $\dots$ &  0.084 &  0.0 &  0.0 &   0.0 &  0.0 &  0.0 &   0.0 & y \\
10:02:33.180$\pm$   0.003 & 2:17:52.76$\pm$      0.00  &  0.461$\pm$  0.129  &   0.376$\pm$  0.148  &  0.087 &  1.7 &  1.5 & -10.3 &  0.0 &  0.0 &   0.0 &   \\
\enddata

\tablecomments{Radio sources with multiple Gaussian fits are flagged
by an 'y' and listed separately in Tab. \ref{tab:radio}. For these
sources no uncertainties in the coordinates and flux densities were
determined (indicated by '$\dots$'). The uncertainty for the absolute
coordinates is of the order of $\sim$ 0.05$^{\prime\prime}$ as derived
from comparison between the results from SAD and SFIND (see also
section \ref{sec:sou}).}

\end{deluxetable}

\clearpage

\begin{deluxetable}{rccccc}
\tablecaption{Radio Source Counts in the VLA-COSMOS Field\label{tab:counts}}
\tablewidth{0pt}
\tablehead{
\colhead{S$_{lower}$} & 
\colhead{S$_{upper}$} & \colhead{$<$S$>$} & 
\colhead{N$_{obs}$} & \colhead{N$_{eff}$} &
\colhead{dN/dS(/S$^{-2.5}$)}
\\
\colhead{[mJy]} & 
\colhead{[mJy]} & \colhead{[mJy]} &  & &
\colhead{[sr$^{-1}$\,Jy$^{1.5}$]}
}
\startdata
0.090 & 0.225 & 0.184 & 41 & 209.0 & 2.78$\pm$1.31 \\
0.225 & 0.319 & 0.279 & 40 &  87.5 & 4.69$\pm$1.07 \\
0.319 & 0.472 & 0.390 & 40 &  62.2 & 4.77$\pm$0.98 \\
0.472 & 0.810 & 0.609 & 41 &  55.3 & 5.86$\pm$1.17 \\
0.810 & 1.660 & 1.090 & 40 &  49.2 & 8.82$\pm$1.94 \\
1.660 & 13.10 & 4.860 & 40 &  44.0 & 24.9$\pm$9.96 \\
\enddata

\tablecomments{Radio sources counts derived for the VLA-COSMOS field
using equations \ref{equ5} to \ref{equ8} to correct for the weighting
and resolution effect. In Fig. \ref{fig:counts}, the comparison to
source counts from other radio deep field survey is shown. The
integrated fluxes were used to derive the Euclidean normalized source counts
dN/dS(/S$^{-2.5}$).}

\end{deluxetable}

\clearpage

\begin{deluxetable}{rccccc}
\tablecaption{VLA-COSMOS Radio Sources with Multiple Gaussian Components\label{tab:radio}}
\tablewidth{0pt}
\tablehead{
\colhead{ID\#} & 
\colhead{R.A.$_{center}$} & \colhead{Dec.$_{center}$} & 
\colhead{R.A.$_{peak}$} & \colhead{Dec.$_{peak}$} &
\colhead{\# of Fits}
}
\startdata
 1  & 09:58:23.271 & 2:08:59.04 & 09:58:23.271 & 2:08:57.63 &  4\\
12  & 09:58:48.492 & 1:58:33.59 & 09:58:48.461 & 1:58:33.59 &  2\\
39  & 09:59:29.293 & 2:28:45.58 & 09:59:29.230 & 2:28:45.11 &  2\\
56  & 09:59:46.350 & 2:36:02.25 & 09:59:46.194 & 2:36:03.19 &  3\\
61  & 09:59:49.756 & 1:56:50.37 & 09:59:49.912 & 1:56:50.37 &  2\\
70  & 09:59:57.931 & 2:18:10.66 & 09:59:57.994 & 2:18:09.72 &  2\\
120 & 10:00:27.534 & 2:21:22.91 & 10:00:27.659 & 2:21:21.03 &  6\\
123 & 10:00:29.478 & 2:41:21.43 & 10:00:29.416 & 2:40:38.66 & 11\\
134 & 10:00:43.146 & 1:46:07.89 & 10:00:43.177 & 1:46:07.89 &  8\\
138 & 10:00:47.631 & 1:59:09.98 & 10:00:47.599 & 1:59:00.58 &  3\\
140 & 10:00:47.602 & 2:09:58.11 & 10:00:47.570 & 2:09:58.58 &  2\\
141 & 10:00:49.134 & 1:49:21.06 & 10:00:49.604 & 1:49:23.88 &  5\\
146 & 10:00:50.074 & 1:49:45.03 & 10:00:50.043 & 1:49:45.97 &  2\\
159 & 10:00:58.163 & 1:51:35.47 & 10:00:58.069 & 1:51:28.89 &  5\\
174 & 10:01:09.303 & 2:17:21.77 & 10:01:09.272 & 2:17:20.83 &  2\\
181 & 10:01:14.059 & 1:54:42.04 & 10:01:13.526 & 1:54:32.64 &  4\\
182 & 10:01:14.877 & 2:02:15.59 & 10:01:14.846 & 2:02:08.54 &  7\\
213 & 10:01:39.263 & 1:51:38.68 & 10:01:39.138 & 1:51:41.03 &  2\\
232 & 10:02:01.198 & 2:13:27.56 & 10:02:01.198 & 2:13:24.27 &  3\\
246 & 10:02:29.947 & 2:09:10.81 & 10:02:29.728 & 2:09:09.87 &  3\\
\enddata

\tablecomments{Radio sources with multiple Gaussian fits which are flagged
in Tab. \ref{tab:list}. We give their identification number (ID) in
Tab. \ref{tab:list}, the (J2000.0) coordinates of the approximate
center of the radio group (R.A.$_{center}$,Dec.$_{center}$), the
(J2000.0) coordinates of the emission peak of the radio group
(R.A.$_{peak}$,Dec.$_{peak}$), as well as the number of Gaussian
fits/components found by SFIND.}

\end{deluxetable}

\end{document}